\documentclass[12pt]{article}
\usepackage{graphics}
\makeatletter
\def\section{\@startsection {section}{1}{\z@}{-3.5ex plus -1ex minus
 -.2ex}{2.3ex plus .2ex}{\large\bf}}
\def\subsection{\@startsection{subsection}{2}{\z@}{-3.25ex plus -1ex 
minus -.2ex}{1.5ex plus .2ex}{\normalsize\bf}}
\makeatother
\makeatletter
\def\theequation{\arabic{section}.\arabic{equation}}
\newcommand{\sect}[1]{\setcounter{equation}{0}\section{#1}}
\@addtoreset{equation}{section}
\renewcommand{\theequation}{\thesection.\arabic{equation}}
\makeatother

\newcommand{\ky}{\kappa_{orb}}
\newcommand{\de}{\partial}

\newcommand{\bra}[1]{\langle{#1}|}
\newcommand{\ket}[1]{|{#1}\rangle}

\newcommand{\nl}{\nonumber \\}
\def\nn{\nonumber}

\def\ba{\begin{array}}
\def\ea{\end{array}}
\renewcommand{\theequation}{\thesection.\arabic{equation}}
\newcommand{\be}{\begin{equation}}
\newcommand{\ee}{\end{equation}}
\newcommand{\bea}{\begin{eqnarray}}
\newcommand{\eea}{\end{eqnarray}}

\renewcommand{\hat}{\widehat}
\renewcommand{\tilde}{\widetilde}
\def\Z{{\bf Z}}

\def\one{\!\!{\hbox{ 1\kern-.8mm l}}}
\newcommand{\hepth}[1]{{\tt hep-th/#1}}
\newcommand{\ex}[1]{{\rm e}^{#1}}
\renewcommand{\a}{\alpha}

 \def\G{\Gamma}

\renewcommand{\thefootnote}{\fnsymbol{footnote}}
\def\one{{\hbox{ 1\kern-.8mm l}}}

\def\ii{{\rm i}}
\def\ee{{\rm e}}

\newlength{\bredde}
\def\slash#1{\settowidth{\bredde}{$#1$}\ifmmode\,\raisebox{.15ex}{/}
\hspace*{-\bredde} #1\else$\,\raisebox{.15ex}{/}\hspace*{-\bredde} #1$\fi}

\begin{document}
\begin{titlepage}
\rightline{DFTT 30/2000}
\rightline{NORDITA-2000/60 HE}
\rightline{NEIP-00-013}
\vskip 1.2cm
\centerline{\Large \bf Is a classical description of stable}
\vskip 0.3cm
\centerline{\Large \bf non-BPS D-branes possible ?
\footnote{Work partially supported by the European Commission TMR
programme ERBFMRX-CT96-0045 and by MURST.}}
\vskip 1.2cm \centerline{\bf M. Bertolini $^a$, P. Di Vecchia $^a$,
M. Frau $^b$, A. Lerda $^{c,b}$, R. Marotta $^a$, R. Russo $^d$}
\vskip .8cm \centerline{\sl $^a$ NORDITA, Blegdamsvej 17, DK-2100
Copenhagen \O, Denmark}
\vskip .4cm \centerline{\sl $^b$ Dipartimento di  Fisica Teorica,
Universit\`a di Torino} \centerline{\sl and I.N.F.N., Sezione di
Torino,  Via P. Giuria 1, I-10125 Torino, Italy}
\vskip .4cm \centerline{\sl $^c$ Dipartimento di Scienze e Tecnologie
Avanzate} \centerline{\sl Universit\`a del Piemonte Orientale, I-15100
Alessandria, Italy}
\vskip .4cm \centerline{\sl $^d$ Institute de Phisique, Universit\'e
de Neuch\^atel} \centerline{\sl Rue A.-L. Breguet 1, CH-2000
Neuch\^atel, Switzerland}
\vskip 0.8cm
\begin{abstract}
We study the classical geometry produced by a stack of stable
(i.e. tachyon free) non-BPS D-branes present in K3 compactifications
of type II string theory. This classical representation is derived by 
solving the equations of motion describing the low-energy dynamics
of the supergravity fields which couple to the non-BPS 
state. Differently from what expected, this configuration 
displays a singular behaviour: the space-time geometry has a 
{\it repulson}-like singularity. This fact suggests that the simplest
setting, namely a set of
coinciding non-interacting D-branes, is not acceptable. We
finally discuss the possible existence of other acceptable configurations
corresponding to more complicated bound states of these non-BPS branes.
\end{abstract}
\end{titlepage}

\renewcommand{\thefootnote}{\arabic{footnote}}
\setcounter{footnote}{0} \setcounter{page}{1} 

\tableofcontents 
\sect{Introduction} 
One of the corner-stones of all recent developments in string theory has
been the exact microscopic description of D-branes   provided by
J. Polchinski \cite{polc95}. D-branes are Ramond-Ramond charged
objects defined as hypersurfaces on which open strings can end. From the
low-energy point of view, the D-branes instead appear as classical
solutions \cite{Hor} of the supergravity field equations which
preserve a fraction of the original set of supercharges, and hence are
BPS saturated.  The fact that type II supergravity possesses classical
BPS  solutions was known well  before Polchinski's paper (see for
instance \cite{dkl95}),
but only after the stringy interpretation their fundamental importance
has been fully appreciated.

More recently, after a series of papers by A. Sen
\cite{Sen1,Sen2,Sen4,Sen5,Sen6},  a lot of attention has been devoted
to the study of non-BPS D-branes (for reviews see
\cite{Senr,leru,schw1,gablec}).  The main motivations are the
following: $i)$ stable non-BPS D-branes are  crucial in testing some
non-perturbative string dualities without relying on supersymmetry
arguments \cite{Sen4,Gab1,Gab2,gallot,gallot1}; $ii)$ the existence of
non-BPS D-branes could be used to define or find duality relations in
a non-supersymmetric context \cite{bergab2}; $iii)$   non-BPS D-branes
hopefully can play a crucial role in describing non-perturbative
properties of non-supersymmetric gauge theories, similarly to what
happened with the BPS D-branes in connection with supersymmetric
Yang-Mills theory. 

Despite the big amount of knowledge that has been accumulated,
non-BPS D-branes have still to be completely understood. There are two
main types of D-branes which do not saturate the BPS bound:  those
which are unstable due to the existence of tachyons on their
world-volume, and those which are instead  stable and free of
tachyons.  For example, the D$p$ branes of type IIA with $p$ odd 
are of the
first kind, while the D-particle of type I is of the second kind. In
both cases,  their microscopic description is fairly well under
control, for instance, in terms of world-volume effective actions
\cite{Senact,russcr} or boundary states
\cite{Sen2,Gab1,gallot,gabstef}~\footnote{For a review of the boundary
state formalism and its applications, see 
\cite{antonella1}.}, but  very little is known about the
nature or even the existence of  the classical geometry associated to
them.  Actually, in the case of an unstable non-BPS brane one should
first of all specify what is the meaning of a classical solution, but even
in the simpler case of the stable non-BPS branes a discussion  about
the conditions which guarantee the consistency of the classical
geometry with the microscopic string description is lacking.

From the effective field theory point of view, the problem of finding
the classical solution corresponding to a given brane configuration
is always well defined, because it amounts to solve the inhomogeneous
field equations of the supergravity theory in the presence of a source
term represented by the brane  effective action, which has a delta-function 
singularity at the position  of the brane.  
Equivalently, the same problem can be 
addressed by solving the
homogeneous field equations and  then imposing that the solution has
the asymptotic behaviour prescribed  by the boundary state description
of the brane.  As we shall see, 
this procedure clearly shows how the integration
constants appearing in  the solution are related to the physical
parameters of the brane  configuration, typically its tension and
charges.  The existence and the nature of possible singularities of the
solution therefore depend on such constants, and
the physical requirements that a classical solution should fulfill 
(for instance, the absence of naked singularities) constrain the 
acceptable range of their values.

From the microscopic point of view, however, the supergravity action
describes only the effective dynamics of the model at low energies,
and thus one has to be sure that the constraints imposed 
on the brane parameters by the existence of a 
meaningful classical solution of the
field equations  are compatible with the approximations 
that lead to such an action.  As is well known, this may
happen only when  the brane tension is very large,
{\it i.e.} $M_p \to \infty$. The
validity  of the no-force condition, which allows to construct a
superposition of an arbitrary large number of D-branes, is therefore
the necessary condition for the consistency of a classical
solution.

In this paper we address these questions in general and discuss in
particular the case of the non-BPS D-particle in six dimensions
arising from the compactification of the type II string theory  on a
K3 manifold at the orbifold point.  This configuration, which can be easily
described using the boundary state formalism, is {\it stable} because the
orbifold projection removes the open string tachyons; moreover, at
a particular value of the volume of the compact space it 
satisfies a {\it no-force condition} at one loop \cite{gabsen}. This
system seems therefore to possess all the required features to produce a
non-trivial classical geometry whose leading behavior
at large distances has recently been found in  \cite{eyras}. 
However, differently from what
expected, we will give evidences that this does not happen.
In particular, we will find that the geometry 
corresponding to a stack of
such non-interacting non-BPS branes displays pathological features
which make the configuration unacceptable. 
This is not in contradiction with the result of~\cite{gabsen}. In fact,
one-loop calculations can extend their validity to the supergravity
regime only when some preserved supersymmetries cancel higher order
corrections. In the case of a non-BPS configuration clearly this is not
guaranteed and our result is indeed an evidence that the no-force
found at first order is lost at two-loop level.
It is interesting to observe that the singularities
we encounter in our solution manifest themselves as 
divergences in the metric tensor
which make the gravitational force to become ``repulsive'' at small
distances. This kind of singularities are known in the literature
as {\it repulsons} \cite{kal}, and have been recently considered
in string theory in \cite{john} where a mechanism for their
resolution has been proposed. It would be interesting to investigate
whether a similar mechanism can be applied also in our case.

The content of this paper is the following: in Section 2 we consider
the unstable non-BPS D-branes of type II in ten dimensions, and discuss 
the limits and the validity of the corresponding classical geometry. 
In Section 3 we compactify the type II string on the
orbifold $T_4/\Z_2$ and write the six-dimensional low-energy effective
action which  describes the model in the field theory limit. Even
though it is known  that this action is that of
the (1,1) supergravity coupled to 20 $U(1)$ vector multiplets,  for our
purposes we find more convenient to recover its explicit form
using the S-duality which relates our model to the
heterotic theory compactified on $T^4$. In Section 4, by exploiting
the precise knowledge of
the action obtained with the duality map, and of the couplings
between the fields and the D-brane given by the boundary state, we
write down the equations of motion and
solve them iteratively in the effective open string coupling $g\, N$.
Contrarily to what expected, the perturbative series may be resummed to
obtain  the exact solution with the asymptotic behaviour described by
the  boundary state.  
As already stressed, the solution presents
pathologies which signal the impossibility of having a macroscopic
stable configuration of $N$ coincident non-BPS D-particles. 
It may be possible that more complicated bound states made up of
stable non-BPS D-branes can resolve these 
singularities and lead to a consistent classical solution. 
This issue is discussed in Section
5, where following \cite{zhouzhu1}, we solve the system of the 
homogeneous field equations in full generality 
and comment on possible different settings in
which a macroscopic solution can exist and the singularity
be resolved. Finally, in Appendix A and B we present some
technical details and explicit calculations.

\vskip 1.5cm

\sect{Non-BPS D-branes in type II theories}
\label{section2}
By now it is well-known that type II string theories in ten dimensions
possess non-BPS D$p$-branes (with $p$ odd in IIA and $p$ even  in
IIB)~\footnote{Throughout this section, we always take $p<7$.}  which
are unstable due to the existence of open string tachyons on their
world-volumes. If we give a vanishing v.e.v. to such tachyons,  these
non-BPS D-branes are easily described by a boundary state $\ket{Dp}$
which has only a NS-NS component
\begin{equation}
\ket{Dp} = \sqrt{2}~ \ket{Bp}_{\rm NS-NS}
\label{bs}
\end{equation}
where $\ket{Bp}_{\rm NS-NS}$ is the GSO projected boundary state in
the NS-NS sector whose explicit expression can be found for example in
\cite{antonella1}. The factor of $\sqrt{2}$ in (\ref{bs})
is required by the open-closed string consistency and implies that a
non-BPS D$p$-brane of type IIA (or B) is  heavier than the
corresponding BPS D$p$-brane of type IIB (or A).  In fact, from
(\ref{bs}) one can see that the tension is
\begin{equation}
\hat{M}_p = \sqrt{2}\, \frac{T_p}{\kappa_{10}\,g}
\label{mp}
\end{equation}
where  $T_p=\sqrt{\pi}\left(2\pi\sqrt{\alpha'}\right)^{3-p}$  is the
factor that appears in the normalization of the boundary state,
$\kappa_{10}=8\pi^{7/2}{\alpha'}^2$ is  the gravitational coupling
constant in ten dimensions, and $g$ is the string coupling constant.

By projecting $\ket{Dp}$ onto perturbative closed string states, one
can easily see that  the only massless bulk fields emitted by the
non-BPS D-branes are the graviton $G_{\mu\nu}$ and the dilaton
$\varphi$, and that their couplings  are described by the DBI action
(in the Einstein frame)~\footnote{ We label the world-volume
directions of the brane with indices $\alpha,\beta,...=0,...,p$ and
the transverse directions by indices $i,j,...=p+1,...,9$.}
\begin{equation}
S_{\rm boundary} = -\hat{M}_p\,\int d^{p+1}x~ {\rm e}^{\frac{p-3}{4}\varphi}
\,\sqrt{-\det{G_{\alpha\beta}}}\quad .
\label{sdbi}
\end{equation} 
The graviton and the dilaton can in principle propagate in the  entire
ten dimensional space-time where their dynamics is governed  by the
following bulk action (again in the Einstein frame)
\begin{equation}
S_{\rm bulk} = \frac{1}{2\kappa_{10}^2} \int
d^{10}x~\sqrt{-\det{G}}\,\left({\cal R}(G) -\frac{1}{2}
\partial_{\mu}\varphi\,\partial^\mu\varphi\right)
\label{sbulk}
\end{equation}
which is a consistent truncation of the type IIA (or B) supergravity
action containing only those fields emitted by the non-BPS D-brane.

\begin{figure}[ht]
\begin{center}
{\scalebox{1}{ \includegraphics{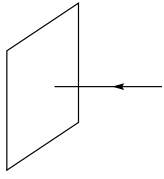} }}
\caption{\small The leading contribution to the one-point function of a
bulk field expressed in a diagrammatic way.}
\label{00}
\end{center}
\end{figure}

Following the procedure  described in \cite{bs,dec1} and using the
explicit form of the boundary state (\ref{bs}), one can find the
metric and dilaton profiles  at large distances from the brane.  These
turn out to be given by
\begin{eqnarray}
G_{\alpha\beta} &\simeq& \left(1+\frac{p-7}{8}\,\frac{\hat{Q}_p}{r^{7-p}} +
...\right)\,\eta_{\alpha\beta} \nonumber \\ G_{ij} &\simeq&
\left(1+\frac{p+1}{8}\,\frac{\hat{Q}_p}{r^{7-p}} +
...\right)\,\delta_{ij}
\label{largedistance0} \\
\varphi &\simeq& \frac{3-p}{4}\,\frac{\hat{Q}_p}{r^{7-p}}+...  \nonumber 
\end{eqnarray}
where $r$ is the distance in transverse space and
\begin{equation}
\hat{Q}_p=\frac{2\,\hat{M}_p\,\kappa_{10}^2\,g^2}{(7-p)\,\Omega_{8-p}}
\label{qp}
\end{equation}
with $\Omega_q=
2\pi^{\frac{1}{2}(q+1)}/\Gamma\Big(\frac{1}{2}(q+1)\Big)$  being the
area of a unit $q$-sphere.  We remark that the expressions
(\ref{largedistance0}) are the same as those of the usual BPS
D$p$-branes in the Einstein frame  (except for the different value of
the tension appearing in $\hat{Q}_p$). It is also worth pointing out that
the terms in (\ref{largedistance0}) proportional to $\hat{Q}_p$ can be
obtained by evaluating the 1-point diagrams for the graviton and the
dilaton (see Figure \ref{00}) in which the couplings with the brane
are read  from the boundary action (\ref{sdbi}).

At this point it is natural to ask whether these non-BPS
D-branes can yield a non-trivial classical geometry in ten dimensions,
just like the BPS D-branes do.  In other words, one can ask whether
the metric in (\ref{largedistance0}) can be interpreted not only as a
small  deformation of the flat Minkowski space-time due to the
emission of a graviton, but also as the asymptotic behavior of a
non-trivial space-time geometry. One way to answer this question is
to compute higher order terms in $\hat{Q}_p$, both for the metric and the
dilaton profiles, and eventually re-sum the perturbative
series. Despite its conceptual simplicity, this procedure clearly
requires calculations which become more and more daunting as one
proceeds in the perturbative expansion.  However, there is another
(and more efficient) way to answer the question, namely one can write
the field equations of the metric and the dilaton  that follow from
the bulk action (\ref{sbulk}), and then look for a solution with the
asymptotic behavior  (\ref{largedistance0}). In our case these
equations are simply
\begin{equation}
\partial_\mu\left(\sqrt{-\det
G}\,G^{\mu\nu}\,\partial_\nu\varphi\right)=0
\label{dil00}
\end{equation}
for the dilaton, and
\begin{equation}
R_{\mu\nu}-\frac{1}{2}\,G_{\mu\nu}\,{\cal R} -\frac{1}{2}
\left(\partial_\mu\varphi\,\partial_\nu\varphi
-\frac{1}{2}\,G_{\mu\nu}\,\partial_\rho\varphi\,\partial^\rho\varphi\right) =
0
\label{metr00}
\end{equation}
for the metric. If we require Poincar\'e invariance in the
world-volume  and rotational invariance in the transverse space, we
can use the following {\it Ansatz} 
\begin{eqnarray}
ds^2 &=& B^2(r)\,\eta_{\alpha\beta}\,dx^\alpha\,dx^\beta  +
F^2(r)\,\delta_{ij}\,dx^i\,dx^j \nonumber \\ \varphi&=& \varphi(r)
\label{ansatz}
\end{eqnarray}
and then solve for the functions $B(r)$, $F(r)$ and
$\varphi(r)$. Using this {\it Ansatz}, the field equations
(\ref{dil00}) and (\ref{metr00}) become
\begin{eqnarray}
&&\varphi'' +\left(\xi' +\frac{8-p}{r} \right)\varphi' = 0  \nonumber
\\ &&(\log{B})'' +\left(\xi' +\frac{8-p}{r} \right)(\log{B})' = 0
\nonumber \\ &&(\log{F})'' +\left(\xi' +\frac{8-p}{r}
\right)(\log{F})' +  \frac{\xi'}{r} = 0 
\label{equations}\\
&&\xi''+(\log{F})'' -\left(\xi' -\frac{8-p}{r}
 \right)(\log{F})'\nonumber \\ &&~~~~~~~~~~~~+ (p+1)\, [(\log{B})']^2
 + (7-p)\, [(\log{F})']^2 + (\varphi')^2 = 0  \nonumber
\end{eqnarray}
where $~'\equiv d/dr$, and
\begin{equation}                
\xi \equiv (p+1)\,\log{B}+ (7-p)\,\log{F} \quad .
\label{xi}
\end{equation}
The general solution of these equations can be easily deduced from the
analysis of \cite{zhouzhu1}~\footnote{See also \cite{oz} for
a related discussion.} and depends on several integration constants
which can be uniquely fixed by imposing the asymptotic behavior
(\ref{largedistance0}) dictated by the boundary state.  If we
introduce the harmonic functions
\begin{equation}
f_\pm(r) = 1\pm\,x\,\frac{\hat{Q}_p}{r^{7-p}}~~,
\label{f+-}
\end{equation}
the non-BPS D-brane solution can be written in a rather simple form
and reads
\begin{eqnarray}
B^2(r) &=& \left(\frac{f_-(r)}{f_+(r)}\right)^\lambda
\nonumber \\ F^2(r)
&=& f_-(r)^{\mu_-}\,f_+(r)^{\mu_+}
\label{solution}\\
{\rm e}^{\varphi(r)} &=& \left(\frac{f_-(r)}{f_+(r)}\right)^\nu \nonumber
\end{eqnarray}
where 
\begin{eqnarray}
\lambda= \frac{7-p}{16\,x}~~~,~~~
\mu_\pm &=& \frac{2}{7-p}\pm \frac{p+1}{16\,x}
~~,~~~\nu= \frac{p-3}{8\,x}~~,
\label{parameters} \\
x&=& \sqrt{\frac{7-p}{8\,(8-p)}} ~~.\nonumber
\end{eqnarray}
It is not difficult to realize that the metric described by
(\ref{solution}) possesses a curvature singularity at
$r_p\equiv(x\,\hat{Q}_p)^{1/(7-p)}$ \cite{oz}. Thus, the solution 
(\ref{solution}) is meaningful only in the ``physical'' region $r>
r_p$.

The problem that we want to address now is the consistency of the
geometrical description (\ref{solution}) with the microscopic string
interpretation of the non-BPS D-branes.  In other words we want to
check whether the classical solution  is consistent with the
approximations that lead to the action from which it descends.
In this respect, we observe that (\ref{sbulk}) is a valid effective
action  only  when curvature effects are small with respect to the
string scale so that higher derivative terms can be consistently
neglected in the Lagrangian.  In our case this  happens when $M_p$ is
large. However, as we have seen at the beginning of this section, the
microscopic interpretation implies that the tension $M_p$ is not a
free parameter, and the only way to make it large is to consider  a
superposition of $N$ D-branes and then take the limit $N\to \infty$.
But this is possible only if the D-branes do not interact  with each
other, {\it i.e.} if they satisfy a no-force condition.

Unfortunately, the unstable non-BPS D-branes in ten dimensions do not
enjoy this property.  To see this, let us compute the interaction
energy $\Gamma$ between two D-branes, which, from the closed string
point of view, is simply given by 
\begin{equation}
\Gamma= \bra{Dp}{\cal P}\ket{Dp}
\end{equation}
where ${\cal P}$ is the closed string propagator. Using standard
techniques, it is easy to see that $\Gamma$ is not vanishing;
moreover, by taking the field theory limit,  one may find that
\begin{equation}
\Gamma\,\Big|_{\alpha'\to 0}=  \hat{M}_p \,V_{p+1}\,\frac{\hat{Q}_p}{r^{7-p}}
\label{force}
\end{equation}
where $V_{p+1}$ is the (infinite) world-volume of the D$p$-brane.
Eq. (\ref{force}) 
explicitly shows that there exists a non-vanishing force between two
non-BPS D-branes: in fact the attraction due to the exchange of
gravitons and dilatons is not compensated by any repulsion because
these branes do not carry any charge.  Thus, according to our previous
discussion, we can conclude that, since it does not satisfy the
no-force condition,  the classical solution (\ref{solution}) is
acceptable (for $r>r_p$)
as long as we do not require a microscopic string
interpretation of the underlying theory. 

One may wonder whether these conclusions may change by taking into
account the presence of tachyons on the brane world-volume.  As is
well-known, these fields   have non-trivial consequences on the
open-string dynamics and modify the structure of the boundary action
\cite{Senact}. Even if these effects are taken into
account, and consequently the form of the solution is changed  (see
for example \cite{oz} for a recent discussion on this point), the
no-force condition still cannot be satisfied.  Furthermore, if we
appeal to the existence of tachyonic modes, a more fundamental
question arises,  namely to what extent a classical geometry can be
associated to an unstable system. 

In conclusion we see that the two essential requirements for the
existence of a consistent geometrical description of a  D-brane are
its stability and the validity of the no-force condition. As we have
seen, these two properties are not satisfied by the non-BPS D-branes
considered so far. However, it turns out that in suitable orbifold
compactifications of type II theories, there exist non-BPS D-branes
which are stable, {\it i.e.}  tachyon free, and do not interact
pairwise, at one-loop, \cite{gabsen}.  
These are therefore the natural candidates to
be considered for a classical supergravity interpretation consistent
with a microscopic string description.  The study of these branes will
be the subject of the remaining part of this paper.
\vskip 1.5cm


\section{Low-energy actions}
\label{section3}

In this section we provide the necessary ingredients to analyze the
geometry associated to the stable non-BPS D-branes in six
dimensions. These are non-perturbative configurations of the type II
string compactified on $T_4/\Z_2$ orbifolds, which have been
extensively studied using the boundary state formalism
\cite{gabsen,gabstef,eyras}.  Here we will focus on the simplest case,
namely the stable non-BPS D-particle. From its boundary state
description it is easy to realize that
such a particle is a source for a graviton, a dilaton, four scalars and
a vector potential in six dimensions. Our goal is to determine the
classical configuration for these fields and study its
consistency. Therefore, in subsection ~\ref{bul} we first derive the
bulk action that governs the dynamics of the fields emitted by the
D-particle. As we will see, this is a consistent truncation of the
$D=6$ supergravity action. Later, in subsection ~\ref{bol} we will
derive the boundary action that describes the couplings of the bulk
fields with the D-particle. 

\vskip 0.7cm
\subsection{Bulk theory}
\label{bul}
The theory we consider is the  non-chiral supergravity in six
dimensions with sixteen supercharges.  To describe it, we start from
the ten dimensional type IIA string compactified on a  torus $T_4$ and
orbifolded with a discrete parity  $\Z_2$ generated by the  reflection
${\cal I}_4$ of the four compact  directions (labeled by indices
$a,b=6,...,9$).  Equivalently, we could start from the type IIB string
compactified on the orbifold generated  by ${\cal I}_4(-1)^{F_L}$,
where $(-1)^{F_L}$ is the operator that changes the sign of all R-R
and R-NS states. These two theories are related to each other  by a
$T$-duality along one of the compact directions, and thus yield the
same low-energy Lagrangian. Both orbifolds have been extensively
studied in the literature (see for example \cite{PolB}) and it is well
known that their massless spectrum is described by the six dimensional
$(1,1)$ supergravity coupled to 20 $U(1)$ vector multiplets. The
action of this theory   is commonly written in the following compact
form \bea \label{sugrad} S & \sim & \int d^6x~ \sqrt{-\det g}~ \ex{-2
\varphi} \Bigg[{\cal R}(g) + 4\,\partial_\mu \varphi\;\partial^\mu
\varphi - {1\over 12}H_{\mu\nu\rho} H^{\mu\nu\rho}  \\  && \nonumber -
{1\over 4} \left(M^{-1}\right)_{IJ} F_{\mu\nu}^I F^{J\;\mu\nu}
+{1\over 8}\, {\rm Tr}\left( \partial_\mu M\,\partial^\mu M^{-1}
\right) + \ldots  \Bigg]~, \eea where $g_{\mu\nu}$ is the string frame
metric, $\varphi$ is the dilaton, $H_{\mu\nu\rho}$ is the field
strength of the NS-NS two-form, and $M$ is the matrix parameterizing
the coset manifold of the scalar fields.  In our case, this coset is
$SO(4,20)/SO(4)\times SO(20)$, which has the right dimension to
accommodate the 80 scalars of 20 vector multiplets. Finally,
$F_{\mu\nu}^I$ contains the field strengths of  all the $U(1)$'s
present in the spectrum and transforms as a vector under $T$-duality 
\footnote{Note that, in general, a $T$-duality transformation mixes
fields which are in different multiplets.}.

The action (\ref{sugrad}) explicitly displays the full $T$-duality
invariance of the theory. However, for our purposes this form is too
general and blurs a few crucial details.  First of all, since our aim
is to  study the theory in the region of its moduli space
corresponding to the orbifold $T_4/\Z_2$, we are not interested in
having a manifestly $T$-dual invariant formulation. Indeed, the
expectation values of the scalars change under $T$-duality, and so
does the shape of the compact space. Secondly, we need to know the
precise normalizations of the various terms in  the action, and in
particular their dependence on the moduli.  In fact, from the
microscopic description we know that the stability of the non-BPS
branes crucially depends on the radii of the compact space
\cite{gabsen}.  Moreover, if we want to construct from these branes a
macroscopic object, the shape of the internal $T_4$ must be further
restricted by fixing all radii to some critical value \cite{Sen6}.
For these reasons, we need to write the supergravity action
(\ref{sugrad}) in a form capable to make more explicit its relation
with the orbifold construction.  This means that we must break the
$SO(4,20)/SO(4)\times SO(20)$ invariance and select, out of the 80
scalars, 4 fields that  describe the characteristic lengths of the
internal space.  As will become clear later, we will also need to know
the relationship between the fields appearing in the supergravity
action and their string counterparts which are associated to   the
massless vertex operators of the orbifold conformal field theory.

It would be interesting to perform these steps directly within the
supergravity context, but this turns out to be a rather non-trivial
task. Therefore, we take a different route and exploit the $S$-duality
that relates our model to the heterotic string compactified on a torus
$T^4$.  In this way, we can carry out the reduction from ten to six
dimensions on the heterotic side by using the standard machinery  of
toroidal compactification,  and then translate the result in the type
II theory by using the duality map. Notice that this short-cut is
possible because this $S$-duality in six dimensions acts trivially on
the  moduli spaces of the two theories.  This means that all couplings
involving the 80 scalars  and their dependence on the internal radii
do not change in going from the heterotic to the type II string.
 
We now sketch the derivation of the type II effective action starting
from the heterotic string compactified on a 4-torus whose low-energy
action (in the string frame) is \bea\label{Sh} S_{\rm h} & =&
{(2\pi\sqrt{\a'})^4\over 2 \kappa_{10}^2} \!\int \!\! d^6 x\,
\sqrt{-\det g^{\rm h}}\,\,\ex{-2 \varphi^{\rm h}}\left[{\cal R}(g^{\rm
h}) +  4\partial_\mu\varphi^{\rm h}\,\partial^\mu\varphi^{\rm h}
+{1\over 4}\partial_\mu \phi_{a}^{\rm h}\, \partial^\mu (\phi_{a}^{\rm
h})^{-1} 
 \right]\nonumber \\  & - & {(2\pi\sqrt{\a'})^4\over 4 g_{10}^2} \int
d^6 x\, \sqrt{-\det g^{\rm h}}~\ex{-2 \varphi^{\rm h}}\;  F^I_{\mu\nu}
F^{I\,  \mu\nu} +\,\dots ~~ \eea where the gravitational and the gauge
couplings are related in the  usual way
$\kappa_{10}/g_{10}=\sqrt{\a'}/2$.  Note that (\ref{Sh}) is only a
subset of the whole action coming from  the toroidal compactification
since  most of the original fields have been  put to zero. We will
discuss the validity of this  truncation after having translated the
action (\ref{Sh}) in the type II language; its motivations will become
clearer when  the boundary action related to the non-BPS brane is
discussed (that  is, when the source term is taken into account).
Here, we just notice that in (\ref{Sh}) only the scalars related  to
dilatations of the compact dimensions have been explicitly written:
since we  consider a compactification where the torus is just a
product  of four orthogonal circles, these scalars are simply the four
diagonal components of the ten-dimensional metric in the internal
space, {\it i.e.}  $g_{aa}^{\rm h}\equiv \phi_a^{\rm h}
$ with $a=6,...,9$, and  the  dilaton $\varphi^{\rm h}$. The v.e.v.'s
of these scalar fields are given by \be\label{defdil0} \langle 
\phi_a^{\rm h}\rangle 
={(R_a^{\rm  h})^2 \over\a'}~~~ \mbox{ and}~~~ \ex{\langle
\varphi^{\rm h}\rangle} = {\a'\over V_{\rm h}^{1/2}} \,\,g' ~,
\end{equation}
where $g'$ is the heterotic string coupling constant, and $V_{\rm
h}\equiv \prod_{a=6}^{9}R_a^{\rm h}$.  In writing (\ref{Sh}) we have
also assumed that the gauge group is broken to $U(1)^{16}$ by suitable
Wilson lines, and $F^I_{\mu\nu}$ denotes the surviving field strengths
($I=1,...,16$).
 
\begin{figure}[ht]
\begin{center}
{\scalebox{.9}{\hspace{-1.5cm}\includegraphics{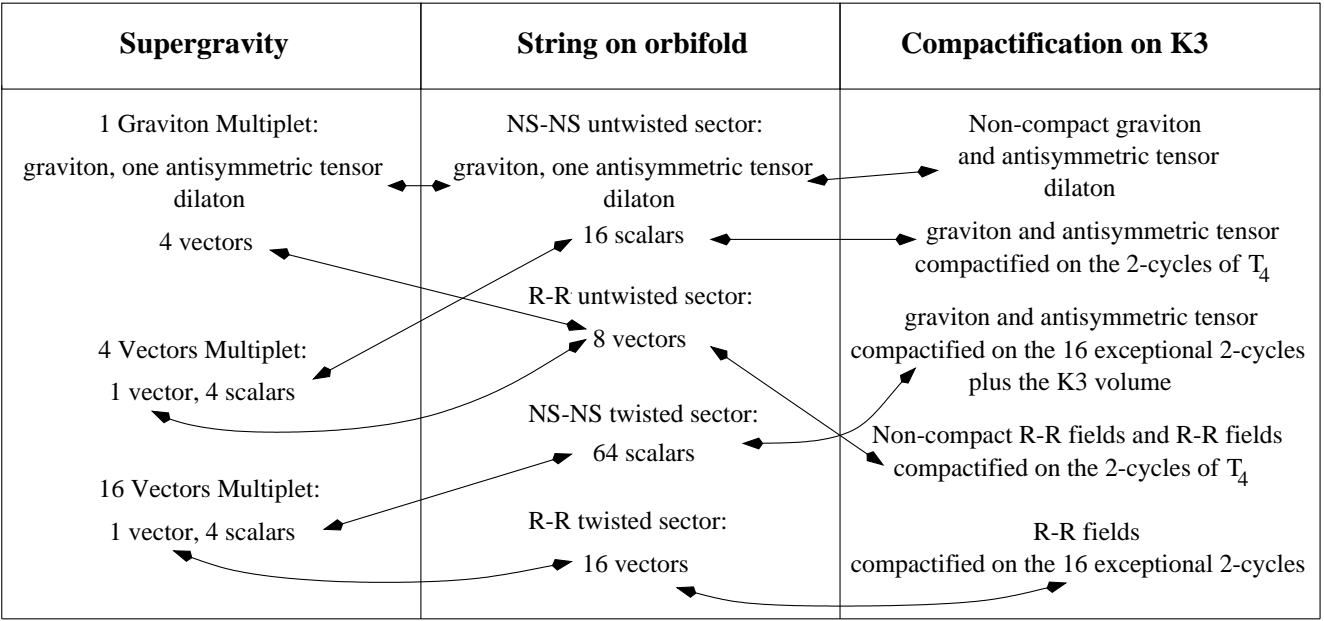}}}
\caption{\small The bosonic low energy spectrum from  three different
points of view}
\label{tab1}
\end{center}
\end{figure}

The correspondence between the heterotic and the type II theories can
be established by means of the following chain of dualities
\be\label{chain}  \hbox{\rm Het. }T^4 
\begin{array}{c} \vspace{-.3cm} S \\ \vspace{.3cm}
\Longleftrightarrow \end{array} \hbox{\rm Type I }T^4 
\begin{array}{c} \vspace{-.25cm} 4\,T \\ \vspace{.25cm}
\Longleftrightarrow \end{array} \hbox{\rm IIB }{T^4\over \Omega {\cal
I}_4}
\begin{array}{c} \vspace{-.25cm} S \\ \vspace{.25cm}
\Longleftrightarrow \end{array} \hbox{\rm IIB }{T^4\over (-1)^{F_L}
{\cal I}_4}  ~\Bigg(\!\begin{array}{c} \vspace{-.25cm} T \\
\vspace{.25cm} \Longleftrightarrow \end{array} \hbox{\rm IIA
}{T^4\over {\cal I}_4}\Bigg)~.
\end{equation}
The last step is not really necessary for our purposes, but it is
useful  to have it in mind. In fact, for some practical calculations
the type IIB picture is easier, whereas the geometrical interpretation
is clearer for  the orbifold of type IIA which is a singular limit of
a smooth K3 manifold.  In Figure~\ref{tab1} we briefly summarize how
the bosonic fields  of this theory emerge from three different points
of view. 

By following the various steps of (\ref{chain}), one can see how the
parameters of the different theories are related to each
other~\cite{PolB}.  For instance, the radii $R^{\rm B}_a $ and the
string coupling constant $g$ in the type IIB  orbifold are related to
the corresponding quantities  of the original heterotic theory as
\be\label{dualB} R^{\rm B}_a = \frac{\sqrt{2}\,V_h^{\,1/2 }}{R^{\rm
h}_a }~~~,~~~  g = \frac{\sqrt{2}\,V_h}{{\alpha'}^2 } \,\,
\frac{1}{g'}~~.
\end{equation}
By performing a further $T$-duality in one of the four compact
directions (say $x^9$) we can reach the type IIA orbifold, for which
we have \be\label{dualA} R^{\rm A}_a=\frac{\sqrt{2}\,V_h^{\,1/2
}}{R^{\rm h}_a }~~\mbox{for} \;a\not=9~~,~~~  R^{\rm A}_9 =
\frac{R^{\rm h}_9\,\alpha'}{\sqrt{2}\,V_h^{\,1/2 }}~~,~~~ g =
\frac{R^{\rm h}_9\,V_h^{\,1/2 }}{{\alpha'}^{3/2 }}\,\, \frac{1}{g'} ~.
\end{equation}
The numerical coefficient in these relations have been fixed by
checking that the masses of BPS objects take the expected values after
a duality transformation. This is the same derivation used in
\cite{Gab2}; however, here we do not follow their conventions and  our
results are slightly different. We keep the string length fixed, {\it
i.e.}  $\a'_{\rm h}=\a'_{\rm B}=\a'_{\rm A}\equiv\a'$, and define  the
dilaton v.e.v. in the orbifold compactification as \be\label{defdil}
\ex{\langle \varphi^{\rm B}\rangle} =  {\sqrt{2}\,\a'\over V_{\rm
B}^{1/2}} \,\,g~,
\end{equation}
$V_{\rm B}=\prod_{a=6}^9 R_a^{\rm B}$, and similarly for the IIA case.
The factor of $\sqrt{2}$ in the above definition is quite natural. In
fact, as usual, the dilaton v.e.v. in a compactified theory  contains
the volume of the compact space. In the toroidal case one has
$\mbox{Vol}\sim V$, while in the orbifold  the $\Z_2$ identification
halves the ``physical'' volume of the internal space: $\mbox{Vol}\sim
V/2 $. 

Recalling that the radii are related  to the v.e.v. of the four scalar
fields $\phi_a$, and using (\ref{defdil0}) and (\ref{defdil}),  we can
lift the above duality maps to the field level and obtain 
\be
\label{dualdil}
\varphi^{\rm A,B} = -\,\varphi^{\rm h}~~~~,~~~~ \phi^{\rm A,B}_a =
2\;{\sqrt{\prod_{b=6}^9 \phi^{\rm h}_b} \over \phi^{\rm h}_a}~~. 
\end{equation}
Finally, by exploiting the invariance of the metric in the Einstein
frame under $S$-duality, one  can find the usual relation between the
string-frame metrics: 
\be
\label{dualmetr}
g_{\mu\nu}^{\rm A,B} = \ex{-2\varphi^{\rm h}}\,g_{\mu\nu}^{\rm h}~~.
\end{equation}
Equipped with this machinery, we are ready to perform the  $S$-duality
on the heterotic action (\ref{Sh}), and rewrite it in terms of IIB
quantities. Using   Eq.s~(\ref{defdil0}), (\ref{dualdil}) and
(\ref{dualmetr}), we get
\bea
\label{SB} 
S_B & =& {(2\pi\sqrt{\a'})^4\over 2 \kappa_{10}^2}
\!\int \!\! d^6 x\, \sqrt{-\det g^{\rm B}}\;\ex{-2 \varphi^{\rm
B}}\left[{\cal R}({g_{\rm B}}) +  4\partial_\mu\varphi^{\rm
B}\,\partial^\mu\varphi^{\rm B}  +{1\over 4}\partial_\mu \phi_a^{\rm
B}\, \partial^\mu (\phi_a^{\rm B} )^{-1} \right] \nonumber \\ & - &
{(2\pi\sqrt{\a'})^4\over 4 g_{10}^2} \int d^6 x\, \sqrt{-\det g^{\rm
B}}\;  F^I_{\mu\nu} F^{I\, \mu\nu} +\,\dots ~.  
\eea
It is not difficult to see that this action is consistent with the
perturbative string amplitudes that can be calculated in the IIB
orbifold.  However, in order to do this comparison one has first to
rewrite the  Lagrangian (\ref{SB}) in the Einstein frame by rescaling
the metric $g_{\mu\nu} = \ex{\tilde{\varphi}} G_{\mu\nu}$.  Here
$\tilde{\varphi} = (\varphi- \varphi_\infty)$, where $\varphi_\infty$
is the constant value of the dilaton at spatial infinity, which in our
case is simply the v.e.v.  defined in (\ref{defdil}). Another
rescaling is usually done on the gauge fields. In fact, in type II
theory these  are taken to be dimensionless regardless of the number
of indices they carry, while on the heterotic side they have canonical
dimensions. Thus, we introduce $\tilde{F}=2\sqrt{\a'}\,g\, F$.
Finally, for later convenience, we write $\phi_a^{\rm B}=
\langle\phi_a^{\rm B} \rangle\,{\rm e}^{2\,\tilde \eta_a^{\rm B}}$. In
terms of these rescaled fields, the action (\ref{SB}) becomes 
\begin{eqnarray} 
S_B &=& {1 \over 2 \kappa_{orb}^2} \!\int \!\! d^6 x\, \sqrt{-\det
G_{\rm B}}\; \Bigg[{\cal R}({G_{\rm B}}) -
\partial_\mu\tilde{\varphi}^{\rm B}\,\partial^\mu \tilde{\varphi}^{\rm
B} - \partial_\mu \tilde\eta_a^{\rm B}\, \partial^\mu
\tilde\eta_a^{\rm B} \nonumber \\  &&~~~~~~~~~~~~~~~~~ - \left.{1\over
4}\left({2 \kappa_{orb}^2\over g_{orb}^2}\right)
\ex{\tilde{\varphi}^{\rm B}}  \tilde{F}^I_{\mu\nu} \tilde{F}^{I\,
\mu\nu}  \right]~,
\label{SB-Ein}
\end{eqnarray}
where  \be
\label{con} \kappa_{orb}^2 = {2 \,\kappa_{10}^2\, g^2\over
(2\pi)^4\, V_{\rm B}}~~,~~ \mbox{and}~~~~~~{2 \kappa_{orb}^2\over
g_{orb}^2}={{\a'}^2\over 4\,V_{\rm B}}~. 
\end{equation}
Contrary to what happens in the heterotic theory, here the
gravitational and gauge couplings have a different dependence on the
radii of the compact space. This fact can be naturally understood   by
comparing the tree-level string amplitudes with the vertices derived
from the field theory Lagrangian. As usual, the moduli dependence of
the couplings is directly related to the ``nature'' of strings
involved in the amplitudes. Since in the heterotic theory both the
gauge and the gravitational fields are made out of the same kind of
closed strings, it is natural that all couplings have a uniform
dependence on the moduli. On the contrary, in the type II setup, the
gauge fields we are looking at come from the twisted R-R sector. In
this case the mode expansion of the string coordinates does contain
momentum along the compact directions. Because of this, the twisted
and untwisted vacua are differently normalized
$$
{}_{U}\langle p_1,n| p_2,m
\rangle_{U}\sim\delta^{(6)}(p_1-p_2)~\delta_{nm} \,V ~~~,~~~
{}_{T}\langle p_1,0| p_2,0 \rangle_{T} \sim \delta^{(6)}(p_1-p_2)~~,
$$
and this difference reflects on the volume dependence of the gauge and
gravitational couplings (\ref{con}). It is interesting to remark that
the gauge kinetic term in (\ref{SB-Ein})  becomes canonically
normalized only for a particular value of the radii of the internal
space, namely at $R_a = \sqrt{\a'/ 2}$.  As we will see in the next
paragraph, this value plays a privileged role also from a different
point of view.

We finally comment on the consistency of the truncation we did in
deriving the effective action (\ref{SB-Ein}).  As we already
mentioned, we have considered only those massless fields which couple
directly to the non-BPS D-particle we want to study, and switched off
all other fields. One may ask whether this truncation is consistent
with the equations of motion of the complete
theory~\cite{stelrev}. In particular, problems can arise if in the
full Lagrangian there are interaction
 terms which are only linear in
one of the fields here disregarded, for instance the twisted NS-NS
scalars, call them $\xi$. In this case, the  equations of motion for
$\xi$ will contain a term which is not automatically vanishing in our
approximation, since it is  independent of the field itself and a
contradiction may arise. However, it can be checked with perturbative
arguments that these terms cannot be present in the complete
Lagrangian. For instance, the twisted NS-NS scalars $\xi$ are
described by vertex operators containing a left and a right spin field
of the internal space. Thus, they have a non-zero $M$-point amplitude,
only if one of the other external vertices also contains these spin
fields. However, this is not the case for the vertices corresponding
to the fields we were considering. This shows that the above mentioned
problems cannot arise with the twisted NS-NS scalars and that we can
safely set them to zero. Similar world-sheet considerations can be
done also for the R-R fields we switched off. For the untwisted
scalars, like the compact off-diagonal entries of ten dimensional
metric $g_{ab}$, it is easier to look directly at the part of the
action where they appear. In fact, as well known, the terms containing
only untwisted fields can be derived by means of the toroidal
compactification from the original ten dimensional description: in the
usual calculation, one has simply to put to zero all the fields which
are odd under the orbifold operation. In this way one can check that the 
untwisted scalars appear at least quadratically in the action, and thus
our truncation is consistent.

\vskip 0.7cm
\subsection{Boundary action}
\label{bol}
{}From the supergravity point of view, the boundary action is seen as
the source term which one must add to the bulk action in order to
describe a D-brane configuration. In string theory, this source can be
efficiently represented by a boundary state and, in particular, by its
overlaps with the massless closed string states.  The boundary states
$\ket{Dp}$ that describe the non-BPS D$p$-branes of the type II
orbifolds,  have been studied in detail in the
literature~\cite{gabsen,gabstef}. Here we just recall the main features that
will be employed in the following section.

A first important point is that $\ket{Dp}$ is non-trivial only in the
NS-NS untwisted and R-R twisted sectors of the
theory~\cite{Gab1,Sen6,gablec}.  In particular, focusing on the
non-BPS D-particle present in the type IIB$/\Z_2$ orbifold, one has
\begin{equation}
\label{D0}
\ket{D0} = \ket{B0}_{{\rm NS-NS},U} + \ket{B0}_{{\rm R-R},T_I}~~,
\end{equation}
where the index $I=1,...,16$ in the twisted part indicates on which
orbifold plane the D-brane is placed. The explicit form of the
coherent states $\ket{B0}$ in (\ref{D0}) and their overlaps with
perturbative closed string states have been studied in~\cite{eyras}.
This  paper, however, uses fields and vertex operators that are
essentially written in the framework of the original  string theory in
ten dimensions. But, if one wants to make contact with the  six
dimensional bulk theory discussed in the previous subsection, one must
use more appropriate fields. These can be easily obtained by
observing, for example, that the vertex operators that describe the
six-dimensional graviton and dilaton have the same structure of their
ten-dimensional analogues, but contain only oscillators with indices
in the non-compact directions.  The internal part of the
ten-dimensional vertices describes instead the six dimensional
scalars.  Keeping this in mind, it is not difficult to find the
relation between the canonically normalized fields $h_{\mu\nu}'$ and
$\varphi'$ used in \cite{eyras} and the canonically normalized fields
$\hat{h}_{\mu\nu}$, $\hat{\eta}_a$, $\hat{\varphi}$ to be used in six
dimensions. This relation is  \bea\label{rele} \hat{h}_{\mu\nu} &=&
h_{\mu\nu}' - \eta_{\mu\nu}  \left[ \frac{\sqrt{2}}{2} ~{\varphi}' -
\frac{1}{4} \sum_a h_{aa}'\right] ~~,~~ \nonumber\\ \hat{\eta}_a &=&
h_{aa}' - \frac{1}{2\sqrt{2}}~ \varphi' ~~,~~ \\ \hat{\varphi} &=&
\frac{3\sqrt{2}}{2}~ \varphi' - \frac{1}{2} \sum_a h_{aa}' ~~.~~
\nonumber \eea Writing the vertex operators associated to the hatted
field in (\ref{rele}) and to the (canonically normalized) twisted R-R
potential $\hat{A}_\mu^I$, we can compute the overlaps with the
boundary state (\ref{D0}) and find the couplings between the bulk
fields and the D-particle. Using the results of \cite{eyras} and the
redefinitions (\ref{rele}), we get 
\bea\label{olap} 
\langle
D0\ket{\hat{h}_{\mu\nu}}  &=& M_0 \,\,V_1\,\kappa_{orb} \,\hat{h}_{00}
~~,~~  \langle D0\ket{\hat{\varphi}} = {M_0\,V_1\over 2}\, \kappa_{orb}\,
\hat{\varphi} ~~~,~~ \\ \nonumber  \langle D0\ket{\hat{\eta}_a} &=&
{M_0\,V_1\over 2}\, \kappa_{orb}\, \hat{\eta}_a ~~~~,~~~ \langle
D0\ket{\hat{A}_\mu^I} =  \sqrt{\frac{8\,V_{\rm B}}{{\alpha'}^2}}\, M_0\,V_1
\,\kappa_{orb}\,\hat{A}_0^I ~~, \eea where
\begin{equation}
M_0=\frac{\sqrt{2}\,T_0}{(2\pi)^2\,\kappa_{orb} \,V_{\rm B}^{1/2}} =
{1\over \sqrt{\a'}\, g}~~ 
\label{mass0}
\end{equation}
is the mass of the non-BPS D-particle and $V_1$ is the (infinite)
length of its world-line. The overlaps (\ref{olap})
represent the one-point functions of the bulk fields encoded in  the
boundary action.  To write it we find convenient to use the same
notation of the previous subsection, and not to work any more with the
canonically normalized hatted fields. The relation between the latter
and the fields appearing in the bulk Lagrangian (\ref{SB-Ein}) is
simply given by 
\begin{equation}
G^{\rm B}_{\mu\nu}= \eta_{\mu\nu} + 2\,\ky\, \hat{h}_{\mu\nu}~~,~~
\hat{\varphi}=\frac{\tilde{\varphi}^{\rm B}}{\ky} ~~,~~
\hat{\eta}_a=\frac{\eta_a^{\rm B}}{\ky}  ~~,~~ \hat{A}_{\mu}^I=
\frac{\tilde{A}_{\mu}^I}{g_{orb}}~,
\label{canf}
\end{equation}
where $\ky$ and $g_{orb}$ are defined in (\ref{con}). Then, it is easy
to realize that the overlaps (\ref{olap}) are consistent with the
following boundary action
\begin{equation}
S_{\rm boundary}=-M_0\int d\tau \,{\rm e}^{-\frac 12
\tilde{\varphi}^{\rm B} - \frac 12 \sum_a \tilde\eta_a^{\rm B}}
\sqrt{-G_{00}^{\rm B}}~+~ M_0\int d\tau\,\tilde{A}_0^I ~.
\label{Sbound}
\end{equation}
Of course, this action does not describe the complete world-volume
dynamics of the non-BPS D-particle. In fact, in deriving it we
considered only the trivial configuration for the fields related to
{\em open} strings and we switched off all non-linear couplings with
{\em closed} strings\footnote{For instance, anomalous couplings,
similar to those of usual D-branes, are present also for non-BPS
branes~\cite{russcr}}. However, for our purposes the action
(\ref{Sbound}) will be sufficient.  Note that its untwisted part can
be obtained also from the action (\ref{sdbi}) of the unstable non-BPS
branes discussed in Section~2:  one has just to perform a toroidal
compactification and remove all fields that are odd under the orbifold
projections. On the other hand, the twisted part of (\ref{Sbound})
accounts for the minimal coupling  of a charged particle with its
gauge field.  Notice that the strength we found for this coupling is
consistent with the $S$-dual interpretation of the D-particle.  In
fact, on the heterotic side this particle corresponds to a
perturbative massive state with  charge $q_{\rm h}=2$ under one of the
16 unbroken $U(1)$'s. Of course, this $U(1)$ charge  does not change
in the duality map and, thus, translating the  heterotic result in
type II language, one should find agreement with the overlap
(\ref{olap}). This is indeed the case; in fact, taking into  account
the relation between the type II gauge fields $\tilde{A}$ and those of
the heterotic theory introduced after Eq. (\ref{SB}), we can see that
the gauge charge $q_{\rm h}=2$ becomes exactly the one that we read
from (\ref{Sbound}). 

So far we have discussed the boundary action for a single non-BPS
brane. However, in order to form a macroscopic object one should
consider a superposition of many branes. Only in this case, in fact,
one can hope that the source creates a smooth classical geometry where
it is possible to neglect string and loop corrections.  Of course, the
dynamics of many  coincident branes is quite complicated and can
radically change the couplings previously derived for a single
object. For BPS configurations, this is not the case because the
D-branes do not interact with each other. Thus the effect of the
superposition of $N$ D-branes is simply to multiply by $N$ the
strength of all couplings. In a non-supersymmetric setup, instead, the
branes in general interact among them in a non-trivial way. However,
as shown in ~\cite{gabsen}, the non-BPS D-particles of the IIB$/\Z_2$
orbifold enjoy two fundamental properties that make them similar to
the usual BPS D-branes. In fact, the orbifold projection always kills
the tachyon zero-mode and, if $R_a^{\rm B} \geq \sqrt{\alpha'/2}$ the
winding excitations of the tachyon, which survive the projection, have a
positive mass$^2$.  Thus, the non-BPS D-particles are stable.
Moreover, if $R_a^{\rm B} =\sqrt{\alpha'/2}$ (which is also the value
where the gauge kinetic term in (\ref{SB-Ein}) becomes canonically
normalized), the ``would-be'' tachyons become massless and an
accidental Bose-Fermi  degeneracy appears in the spectrum. Thus, at
the critical radii the force between two non-BPS D-particles vanishes
at one-loop \cite{gabsen},  {\it i.e.}
\begin{equation}
\label{0force}
\Gamma =\bra{D0}{\cal P}\ket{D0} = 0~.
\end{equation}

For this reason, it is natural to conjecture that, in this particular
case, it is possible to describe $N$ non-BPS D-particles simply by
taking the na\"{\i}ve sum of $N$ boundary states previously introduced
to describe a single object, that  is
\begin{equation}
\label{NB}
\ket{D0,N} = N \ket{D0}~.
\end{equation}
We want to stress that the no-force condition  (\ref{0force}) is
clearly a necessary ingredient for this simplification to hold, but it
is not sufficient to really prove the validity of the assumption
(\ref{NB}). In fact, Eq. (\ref{0force}) proves the vanishing of the
interaction only at one-loop (from the open-string point of view), and
does not guarantee that a similar result occurs at higher loops. In
other words, from Eq. (\ref{0force}) we can see that the interactions
between $N$ D-particles vanish at the leading order in $N$, while we
know that the supergravity description is reliable in the opposite
regime, $N\to \infty$. Thus, we take Eq. (\ref{NB}) as a working
hypothesis and, in the next section, we will check whether this leads
to acceptable space-time configurations for the metric and the other
fields.

\vskip 1.5cm
\section{The non-BPS D-particle solution}
\label{section4}
In the previous section we have shown that the action describing the
dynamics of the fields emitted by a non-BPS D-particle in six
dimensions, is given by the sum of Eq.s (\ref{SB-Ein}) and
(\ref{Sbound}) which we rewrite here in a simplified notation 
\bea
\label{sum1}
S&=&\frac{1}{2\kappa_{orb}^2}\int d^6x \,\sqrt{-\det G}\left[{\cal
R}(G) - \partial_\mu\varphi\,\partial^\mu\varphi - \partial_\mu\eta_a
\,\partial^\mu\eta_a - \frac{1}{4}\,  {\rm
e}^{\varphi}\,F_{\mu\nu}F^{\mu\nu}\right]  \nn \\ &&-\,M  \int d\tau
\,{\rm e}^{-\frac 12 \varphi - \frac 12 \sum_a \eta_a}  \sqrt{-G_{00}}
\,+ \,M\int d\tau \,A_0 ~~.\eea 
Notice that here we have fixed the
compact volume to its critical value $V_c={\alpha'}^{\,2}/4$ where the
no-force condition holds  at one loop, and, according to our working
hypothesis (\ref{NB}), we have put $M=N\,M_0$.  The field equations
that follow from this action are
\footnote{We use the static gauge $X^0=\tau$.}
\begin{equation}
\frac{1}{ \sqrt{-\det G}}
\partial_\mu\left( \sqrt{-\det
G}\,G^{\mu\nu}\,\partial_\nu\varphi\right) - \frac{1}{8}\,{\rm
e}^{\varphi}\,F_{\mu\nu}F^{\mu\nu}
= \frac{1}{2}\,T(x)\,\delta^5(\vec
x)
\label{dil1}
\end{equation}
for the dilaton,
\begin{equation}
\frac{1}{ \sqrt{-\det G}}\,\partial_\mu\left( \sqrt{-\det
G}\,G^{\mu\nu}\,\partial_\nu\eta_a\right) =
\frac{1}{2}\,T(x)\,\delta^5(\vec x)
\label{scala1}
\end{equation}
for the 4 scalar fields,
\begin{equation}
\partial_\mu\left( \sqrt{-\det G}\,
{\rm e}^\varphi\, F^{\mu\nu}\right) =  -
2\, M \,\kappa_{orb}^2  G^\nu_{\;\,0}\,\delta^5(\vec x)
\label{vec1}
\end{equation}
for the gauge field, and
\[
R_{\mu\nu} - \frac{1}{2}\, G_{\mu\nu}\, {\cal R} -  \left(
\partial_\mu\varphi\,\partial_\nu\varphi -\frac{1}{2}\,
G_{\mu\nu}\,\partial_\rho\varphi\,\partial^\rho\varphi \right) -
\left( \partial_\mu\eta_a\,\partial_\nu\eta_a -
\frac{1}{2}\,G_{\mu\nu}\,\partial_\rho\eta_a\,\partial^\rho\eta_a
\right)
\]
\begin{equation}
- \frac{1}{2} \,{\rm e}^\varphi \left( F_{\mu\rho}\,F_{\;\,\nu}^{\rho} -
\frac{1}{4}  \,G_{\mu\nu}\, F_{\rho\sigma}F^{\rho\sigma} \right)=
G_{00}\,G_\mu^{\;\,0} G_\nu^{\;\,0}\,T(x)\, \delta^5(\vec{x})
\label{met1}
\end{equation}
for the metric, where
\begin{equation}
\label{tx}
T(x) = -\,M\,\kappa_{orb}^2\,{\rm e}^{-\frac 12  \varphi - \frac 12
\sum_a  \eta_a} \,\frac{\sqrt{- G_{00}}}{\sqrt{-\det G}}
\end{equation}

Our task is to find a solution to these equations describing a static
and spherically symmetric non-BPS D-particle in which the fields
depend only on the distance in transverse space, $r$. There are
several ways to reach this goal. A first possibility is to build up
the solution iteratively via a perturbative approach by expressing
the various fields as series in powers of $1/r^3$ with
arbitrary coefficients (recall that the usual $1/r^{D-p-3}$ dependence
of a $p$-brane in $D$ dimensions 
reduces in the present case to $1/r^3$). Inserting this {\it Ansatz} in the
coupled system (\ref{dil1})--(\ref{met1}), one can determine all
coefficients by solving the equations order by order in $1/r^3$.  
For example, up to third order in $1/r^3$ the  solution
looks like   \bea  \varphi&\simeq& \frac{1}{4}\,\frac{Q}{r^3} -
\frac{1}{8}\left(\frac{Q}{r^3}\right)^2 + \frac{1}{24}
\left(\frac{Q}{r^3}\right)^3 + ... 
\label{dil20} \\
\eta_a&\simeq& \frac{1}{4}\,\frac{Q}{r^3}+... \label{scala20} \\
A_0&\simeq&-\frac{Q}{r^3} + \frac{1}{2}\left(\frac{Q}{r^3}\right)^2  -
\frac{1}{3} \left(\frac{Q}{r^3}\right)^3 
\label{vec20} + ...\\
G_{00}&\simeq& -1 + \frac{3}{4}\,\frac{Q}{r^3} -
\frac{21}{32}\left(\frac{Q}{r^3}\right)^2 + \frac{61}{128}
\left(\frac{Q}{r^3}\right)^3 + ... \label{met200} \\
G_{ij}&\simeq&\delta_{ij} \left[1 + \frac{1}{4}\,\frac{Q}{r^3} -
\frac{3}{32}\left(\frac{Q}{r^3}\right)^2 + \frac{5}{384}
\left(\frac{Q}{r^3}\right)^3  + ...\right]\;,
\label{met20}
\eea 
where
\begin{equation}
\label{Q0}
Q=\frac{2\, M\,\kappa^2_{orb}}{3 \,\Omega_4} \sim
N\,g\,{\alpha'}^{\,3/2} 
\end{equation}
and the indices $i,j=1,...,5$ label the transverse directions. 

\begin{figure}[ht]
\begin{center}
{\scalebox{1}{\hspace{-2cm}\includegraphics{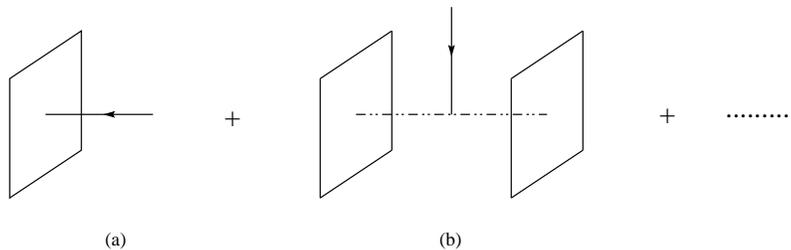} }}
\caption{\small The leading contribution to the one-point function of
a bulk field expressed in a diagrammatic way.  Diagram (a) yields the
leading term in the large distance expansion and is proportional to
$Q$, whereas  diagram (b) corresponds to the next-to-leading
correction proportional  to $Q^2$.}
\label{dil}
\end{center}
\end{figure}

This same result can be obtained via an alternative approach based 
on the use of the boundary state, which, as shown in  \cite{bs,dec1},
allows to find the asymptotic behavior of the various fields at large
distance from the source. For the non-BPS D-particle, this method
has been recently used in \cite{eyras} where the leading terms
proportional to $Q/r^3$ have been obtained, and generalized for any $p$ in
\cite{loz} (provided the redefinitions (\ref{rele}) are taken into account). 
Actually, one can do
more.  Remembering that $Q$ is proportional to the 't Hooft coupling
$\lambda \sim N g$ (see Eq. (\ref{Q0})),  the above expansions for
large $r$ can also be interpreted as expansions  in $\lambda$.  Since
in string theory different powers of $\lambda$ characterize open
string diagrams of different topologies, the various terms in
(\ref{dil20})-(\ref{met20}) can be associated to the one-point
functions of massless bulk fields evaluated on world-sheets with an
increasing number of boundaries. Specifically, the terms linear in $Q$
arise from one-point function on a disk diagram,  the terms
proportional to $Q^2$ from one-point functions on an annulus, and so
on. However, for the purpose of finding the classical solution, it is
not really necessary to perform such calculations in string theory,
but it is sufficient to do them directly in the low-energy field
theory described by the action (\ref{sum1}).  Here one has simply to
compute (in configuration space) diagrams like the ones represented in
Figure \ref{dil}, where the couplings with the sources are determined
by the boundary part of the effective action and the interaction
vertices from its bulk part.
Despite its conceptual simplicity, this diagrammatic method requires
calculations which become more and more cumbersome as one proceeds in
the perturbative expansion. Nevertheless, it is useful because it
clarifies the origin and the meaning of the various terms. In Appendix
A we present  the detailed calculations of the diagrams that
contribute to the classical solution up to $Q^2$.

Let us now return to the perturbative expansions
(\ref{dil20})-(\ref{met20})  of the D-particle solution. Differently
from what expected \cite{eyras}, it turns out that it is possible to
re-sum these series and present  the fields in a closed form. Indeed we
find   \bea
\varphi&=&\frac{1}{4}\,\ln\left[1+\sin\left(\frac{Q}{r^3}\right)\right]
\label{dil3} \\
\eta_a&=&\frac{1}{4}\,\frac{Q}{r^3} 
\label{scala3} \\
A_0 &=& -1 +
\frac{\cos\left(\frac{Q}{r^3}\right)}{1+\sin\left(\frac{Q}{r^3}\right)}
\label{vec3} \\
G_{00}&=& - \left[1+\sin\left(\frac{Q}{r^3}\right)\right]^{-3/4} 
\label{met30} \\
G_{ij}&=& \delta_{ij}
\left[1+\sin\left(\frac{Q}{r^3}\right)\right]^{1/4}~~.
\label{met3}
\eea It is not difficult, but rather tedious, to check that
(\ref{dil3})-(\ref{met3}) is indeed a solution of the differential
equations (\ref{dil1})-(\ref{met1}).  In Appendix B we will provide
all details for deriving the above expressions directly from the field
equations (\ref{dil1})-(\ref{met1}), and present also  their
extension to the case of a generic $p$-brane.

We now discuss the properties of the solution
(\ref{dil3})-(\ref{met3}).  First of all, we observe that it is
well-defined only for $r\geq Q^{1/3}$. In fact, in the region
$r<Q^{1/3}$  the dilaton, the gauge field and $G_{00}$ have branch cut
singularities at  \bea
\label{rinf}
r_n=\left[\frac{Q}{(3/2+ 2n)\pi}\right]^{1/3} \qquad {\rm for} \quad
n=0,1,2,... ~~.  \eea Moreover, the scalar curvature ${\cal R}$ diverges
at these  singular points.   Clearly, when the curvature is big,   the
classical supergravity description is not any more reliable; moreover, since
the singularities at $r=r_n$ are naked, the entire solution is
unacceptable according to the cosmic censorship  conjecture.  This
result seems to indicate that the  prediction made in \cite{gabsen},
about the possibility of having  a classical description  for stable
non-BPS D-branes, does not hold, at least in the case that we have
considered here.  Let us be more specific about this fact.  From the
supergravity point of view, $Q$ is a  free parameter, and solutions
with different  values of $Q$ are all on equal footing. Hence they are
all unacceptable for the reason we have explained above.  However, if
we appeal to the underlying string theory, some crucial  differences
come into play. First of all,  $Q$ is not any more a free parameter
since it is related to the fundamental quantities of the microscopic
theory as shown in (\ref{Q0}).  Moreover, in a string context one may
expect {\it a priori} that the classical  supergravity description can
break down at distances of the order of the string scale, where the
massless closed string states cease to be good probes for the
geometry. Thus, at $r\sim\sqrt{\alpha'}$ stringy effects must be taken
into account, and the entire supergravity approximation must be
reconsidered. In other words, if the naked singularities of a
classical solution are at distances smaller or equal to the string
scale, no contradictions arise, and the supergravity solution can be
accepted at larger distances.  
 
This is what happens for a single D-particle, {\it i.e.} $N=1$.  In
fact, since the first  singularity is at
$r_0=(2Q/3\pi)^{1/3}<Q^{1/3}$, and, for $N=1$, $Q^{1/3}\sim g^{1/3}
\sqrt{\alpha'}$, the region where the classical solution starts to
have problems is inside the region in which stringy corrections  are
relevant. Thus, Eq.s (\ref{dil3})-(\ref{met3}) represent a valid
classical solution associated to a stable non-BPS D-particle  in six
dimensions at distances much larger than the string scale.   However,
if we also want to justify the classical  approximation and give a
reason for disregarding loop  corrections in the bulk, we must also to
take the limit $g\to 0$.  In this case the fields produced by the
D-brane become just small fluctuations around the trivial background
and the only relevant terms are the leading ones in the $Q/r^3$
expansion.

Things are different for $N\not= 1$. In fact, as is clear from
(\ref{Q0}), if we increase the value of  $N$, the parameter $Q$
becomes macroscopic ({\it i.e.}  $Q^{1/3}\gg\sqrt{\alpha'}$), and the
solution (\ref{dil3})-(\ref{met3}) exhibits naked singularities  also
in a region which is not affected by stringy fuzziness effects and
where the classical approximation is reliable.  Therefore, according
to the cosmic censorship conjecture, the solution
(\ref{dil3})-(\ref{met3}) must be rejected, and its  source, namely a
stack of many {\it non-interacting} D-particles, must be regarded as
non physical. 

We would like to stress that our conclusions are not in contradiction
with the result of \cite{gabsen} about the vanishing of the force
between two non-BPS D-particles at critical radius.  In fact, the
result of \cite{gabsen} is exact in $\alpha'$  but perturbative in the
't Hooft coupling $\lambda \sim N g$,  and is due to a cancellation
occurring at one loop because of an accidental Bose-Fermi degeneracy
of the open string spectrum.  The classical solution
(\ref{dil3})-(\ref{met3}) is instead valid in a very different regime,
since it is perturbative in  $\alpha'$ but exact in
$\lambda$. Therefore, our result should be compared with that of
\cite{gabsen} only in the limit $\lambda\to 0$, and if we do this, we
too find a vanishing force at the first order in $\lambda$.  This can
be easily seen by inserting the classical solution
(\ref{dil3})-(\ref{met3}) into the boundary part of action
(\ref{sum1}).  Expanding at first order in $Q$, and subtracting the
vacuum energy, we find
\begin{eqnarray}
S_{\rm boundary}&=&
- M \int d\tau \,\,{\rm e}^{-\frac 12 \varphi
 - \frac 12 \sum_a \eta_a
} \,\, \sqrt{-G_{00}
} + M \int d\tau \,A_0
\nn \\ &\sim&  - M \int d\tau \,\,\frac{Q}{r^3}\,\left(-\frac{1}{8} -
\frac{1}{2} - \frac{3}{8}  + 1\right) = 0~~.
\label{bound3}
\end{eqnarray}
A similar calculation shows however, that the no-force condition is
{\it not} satisfied at the next-to-leading order.  It would be
interesting to derive this  result also from an open string
computation at two loops (see \cite{lamb1} for a recent discussion on
this issue).  In this case the non-trivial dynamics of the open
strings living on the non-BPS D-brane becomes relevant, and due to the
lack of supersymmetry one does not expect any cancellation to occur.  

\begin{figure}[ht]
\begin{center}
{\scalebox{.9}{\hspace{0cm}\includegraphics{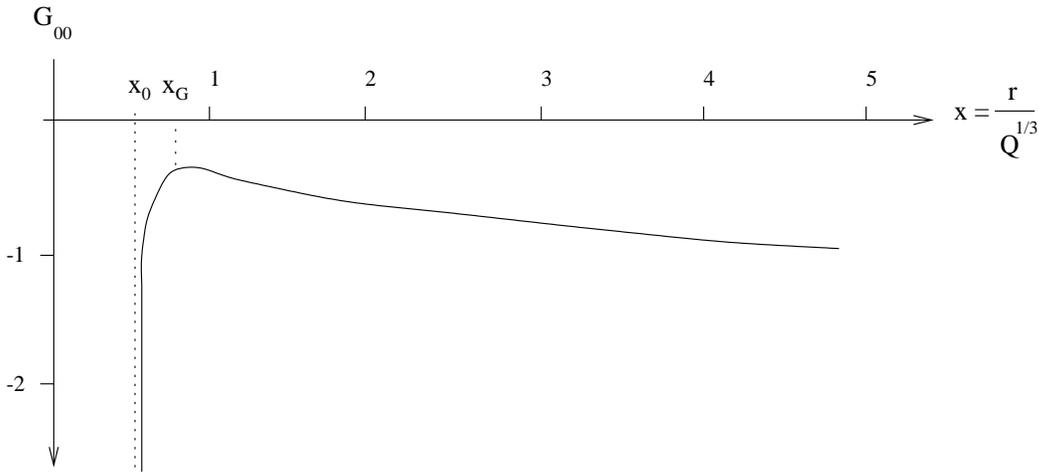}}}
\caption{ \small This is the behavior of $G_{00}$ in the region around $x
\equiv r/Q^{1/3}=1$.
The derivative of $G_{00}$ changes sign around
$x_G\sim 0.8$,  while the first singularity (the repulson) is at
$x_0\sim 0.6$.}
\label{g00}
\end{center}
\end{figure}

We conclude this section with a final observation. The fact that
the stable non-BPS D-particles do not satisfy a
no-force condition at all orders in $\lambda$, is also suggested
by the behavior of $G_{00}$ around $r\sim Q^{1/3}$.  One can check that,
before reaching the first singularity at $r_0$, the derivative of
$G_{00}$ changes sign in $r=r_G > r_0$, see Figure \ref{g00}.
As is well known, this fact indicates that something strange is
happening: indeed, the gravitational force changes its sign at $r=r_G$, and the
singularity located at $r=r_0$ (where $G_{00} \rightarrow - \infty$) 
can be called {\em repulson}~\cite{kal}, because close to it, 
massive particles 
repel each other. These naked singularities have been recently studied
in the context of  string theory in~\cite{john} where a mechanism for their
resolution has been proposed. Since our configuration exhibits properties
similar to those discussed in~\cite{john}, it would be interesting to
see whether the same kind of mechanism works also in our case and resolves
the singularity we have found.

\vskip 1.5cm
\section{General solution and discussion}
\label{section5}

As we have stressed several times, the explicit form of the  solution
(\ref{dil3})-(\ref{met3}) crucially depends on the detailed knowledge
of the source term, and in particular on the hypothesis (\ref{NB}).
In this last section, we relax this assumption and consider a more
general solution of the field equations (\ref{dil1})-(\ref{met1}).  In
fact, on general grounds, one may expect that there exist more
complicated  configurations of non-BPS D-particles  which are stable
and do not display any pathological behavior in the corresponding
classical geometry.  For example, one can think of a non-trivial bound
state of non-BPS D-particles described by some complicated boundary
state $|B\rangle$ and, correspondingly,  by a  boundary action which
could differ from  (\ref{Sbound}) even in its functional dependence on
the fields.  

To explore these possibilities, we therefore study the general
solution  of the differential equations (\ref{dil1})-(\ref{met1})
under the minimal amount of requirements. We simply ask that the
solution describe an asymptotically flat geometry, be spherically
symmetric in  the transverse directions and also that all scalar
fields have vanishing v.e.v. at infinity. Instead, we do not enforce
any specific behavior on the leading terms in the large distance
expansion.  In fact, as we have explicitly seen in the previous
section, these are directly related to the specific microscopic
structure of the source.  Our only hypotheses about it are therefore
that it couples to the graviton, the dilaton, the scalars $\eta_a$ and
to one twisted R-R gauge potential. As we have shown in Section 3,
this set of fields defines a consistent truncation  of the full
six-dimensional supergravity theory. 

Under these assumptions, we now study the most general solution of the
field equations (\ref{dil1})-(\ref{met1}), after removing the source
terms in the right hand sides.  The resulting homogeneous equations
can be analyzed by generalizing the methods of
\cite{zhouzhu1} and the  general solution can be written in a
closed form in terms of elementary functions. It will depends on some
integration constants (two for each equation); half of them are fixed
by the general requirements discussed above, and the remaining ones
are free parameters which can be associated to the microscopic
structure of the source.  In Appendix B we will provide the details to
solve explicitely the homogeneous field equations
(\ref{dil1})-(\ref{met1}); here we simply write the result.  To do
this, it is first convenient to introduce the functions \bea
\label{f300} 
f_{\pm}(r) &=& 1 \pm x\, \frac{Q}{r^3} \\ X(r) &=& \alpha + \beta \ln
 \left(\frac{f_{-}(r)}{f_{+}(r)} \right) \nn \eea  where $Q$ is
 defined in (\ref{Q0}), and $\alpha,\beta$ and $x$ are
 constants. Then, the general solution  gets the following form  \bea
\label{solgen} 
e^{\eta_a} &=& \left(\frac{f_{-}(r)}{f_{+}(r)}\right)^{\delta}  \\
\label{solgen1} 
e^{2\varphi} &=& \left(\frac{\cosh X(r) + \gamma \sinh X(r)}{\cosh
\alpha + \gamma \sinh \alpha} \right)
\left(\frac{f_{-}(r)}{f_{+}(r)}\right)^{\frac{3}{4}\epsilon} \\
\label{solgen2}  
A_0 &=&  \sqrt{2(\gamma^2 - 1)} \left(\frac{\sinh X(r) \left(\cosh
\alpha + \gamma \sinh \alpha\right) }{\cosh X(r) + \gamma \sinh X(r)}
- \sinh \alpha \right) \\
\label{solgen3} 
G_{00} &=& - \left(\frac{\cosh X(r) + \gamma \sinh X(r)} {\cosh \alpha
+ \gamma \sinh \alpha}\right)^{-\frac{3}{2}}
\left(\frac{f_{-}(r)}{f_{+}(r)}\right)^{\frac{3}{8}\epsilon} \\ 
\label{solgen4} 
G_{ij} &=& \delta_{ij} \left(\frac{\cosh X(r) + \gamma \sinh
X(r)}{\cosh \alpha + \gamma \sinh \alpha}\right)^{\frac{1}{2}}
\left(f_{-}(r)\right)^{\frac{2}{3} - \frac{1}{8}\epsilon}
\left(f_{+}(r)\right)^{\frac{2}{3} + \frac{1}{8}\epsilon}   \eea
where  
\begin{equation}  
\epsilon = \pm \frac{4}{3}\sqrt{4 - 3 \beta^2 - 12 \delta^2} ~~.
\label{gensol}
\end{equation}
As anticipated, this solution depends on some integration constants,
namely $\alpha$, $\beta$, $\gamma$, $\delta,$ and $x$, which can be
fixed by  specifying the form of the source term.   For example, if we
impose  the boundary conditions corresponding to $N$ coincident and
non-interacting D-particles, which was the physical situation
considered in Section 4, namely if we require that at large distance
the fields behave as  
\bea
\label{dil2} 
\varphi\simeq \frac{1}{4}\,\frac{Q}{r^3} +...~~,~~~~ \eta_a\simeq
\frac{1}{4}\,\frac{Q}{r^3} +...~~,~~~~  A_0\simeq -\,\frac{Q}{r^3}
+...~~,\\ G_{00}\simeq  -1+\frac{3}{4}\,\frac{Q}{r^3} +...~~,~~~~
G_{ij}\simeq\delta_{ij} \left(1 + \frac{1}{4}\,\frac{Q}{r^3} +
...\right)\;, \nn  
\eea  
one finds that the constants 
$\beta$, $\gamma$, $\delta,$ and $x$ must be chosen as  
\bea
\label{parfix}
x \,\beta= \pm\frac{\rm i}{4} \quad,\quad 
\gamma= \frac{-\sinh 2\alpha \pm\ii}{\cosh 2\alpha} \quad,\quad  x\, \delta=
-\frac{1}{8} \quad   {\rm and}\quad x\to 0  
\eea 
Inserting these
values into  (\ref{solgen})-(\ref{solgen4}), one can easily check that
the solution given in  Eq.s~(\ref{dil3})-(\ref{met3}) is recovered and
the $\alpha$-dependence drops out.
Notice, in particular, that even if the vanishing value for $x$
renders both $f_-$ and $f_+$ trivial, the function $X(r)$  and the
scalars $\eta_a$ do not become constant since the products $x\,\beta$
and $x\,\delta$ are non zero. On the other hand, the fact that 
$x\,\beta$ is purely imaginary, 
makes all hyperbolic  functions become periodic.

The structure of the general solution (\ref{solgen})-(\ref{solgen4})
clearly indicates that different choices of the integration constants
do not  necessarily yield a classical geometry  with a pathological
behavior.  However, when one regards the supergravity as the
low-energy description of string theory, one should ask which of all
possible choices in the classical context have some physical
interpretation from the  string viewpoint. A first very natural
possibility is to consider an orbifold compactification with an
internal volume $V\not=V_c$ (recall that at $V_c={\alpha'}^2/4$  the
non-BPS D-particle becomes ``extremal'', in some sense). In fact,
provided that $V>V_c$, the D-particle remains a stable configuration
\cite{Sen6}, even if it does not exhibit any more the Fermi-Bose
degeneracy which, at first order, was responsible for the no-force
condition. However, since this  condition does not hold at higher
orders, it is not necessary to focus on the critical volume any more,
and one can hope that the departure from extremality will eventually
lead to develop an event horizon which, hiding all singularities in a
casual disconnected region from the physical space, would make the
solution (\ref{solgen})-(\ref{solgen4}) free of singularities.

Another possibility could be to consider a stable bound state
obtained by displacing along the transverse directions the stack of
D-branes in a sphere of radius $r \sim Q^{1/3}$. This kind of
mechanism for resolving the repulsive singularities which we have
found in our solution, is similar to the one advocated in \cite{john},
where these problems have been discussed in great detail. In our
non-BPS situation, this option, however,  deserves further
investigation.  Of course, one could also imagine more exotic
possibilities  that rely on very different settings, with different
source terms and more bulk fields that couple to them. These
configurations would need a different truncation of the original
$(1,1)$ supergravity theory discussed in Section 3, and are clearly
not accomplished by the solution (\ref{solgen})-(\ref{solgen4}). 

The highly non-trivial role that stable non-BPS D-branes could play in
deeper understanding of non-perturbative dualities in string theory
and eventually on non-supersymmetric versions of the AdS/CFT
correspondence, clearly makes quite challenging to find some positive
answers to these open problems.

\vskip 1.5cm {\large {\bf Acknowledgments}}

\noindent
We would like to thank L. Andrianopoli, M. Bill\`o, L. Gallot,
A. Liccardo, I. Pesando and M. Trigiante for very useful
discussions. M.F., A.L. and R.R. thank NORDITA, and M.B. the Physics
Institute of the University of  Neuch\^atel for kind
hospitality. M.B. acknowledges support by INFN and  R.R. by the Fond
National Suisse. 

\vskip 1cm
\noindent
\appendix{\large{\bf {Appendix A}}}

\renewcommand{\theequation}{A.\arabic{equation}}
\setcounter{equation}{0}
\vskip 0.5cm
\noindent
In this appendix we present the diagrammatic calculations of the
leading and next-to-leading terms in the large distance expansion of
the fields emitted by a non-BPS D-particle.  To do this, we first
rewrite the bulk part of the action (\ref{sum1}) in terms of canonical
normalized fields (see Eq. (\ref{canf})) and get
\begin{equation}
S_{\rm bulk} = \int d^6x \,\sqrt{-\det G}
\left[\frac{1}{2\kappa_{orb}^2}{\cal R}(G) -
\frac{1}{2}\partial_{\mu}\hat \varphi \partial^{\mu}\hat \varphi -
\frac{1}{2}\partial_{\mu} \hat \eta_a \partial^{\mu}\hat \eta_a -
\frac{1}{4}  {\rm e}^{\kappa_{orb} \hat \varphi}\,\hat F^{\,2}\right] 
\label{bulk2}
\end{equation}
with $G_{\mu\nu}= \eta_{\mu\nu} + 2\,\ky\, {\hat h}_{\mu\nu}$.
Expanding (\ref{bulk2}) in $\ky$, we get
\begin{equation}
S_{\rm bulk}=S_0 + \kappa_{orb}\, S_I + {\cal O}(\kappa_{orb}^2)
\label{action}
\end{equation}
where $S_0$ is the free action and 
\begin{equation}
S_I= S_{\varphi\varphi  h}+  S_{A A h}+   S_{\eta \eta h} +  S_{h h h}~~.
\label{intac}
\end{equation}
The four terms in $S_I$ describe respectively  the interaction of a
graviton with two  dilatons, of a graviton with two gauge fields,  of
a graviton with two scalars, and  the coupling among three gravitons.  

The interactions of the bulk fields with the D-brane are encoded in
the boundary part of the action (\ref{sum1}), which,  at the
linearized level, is
\begin{equation}
S_{\rm boundary}=\int d^6 x \left( J^{\mu\nu}\, {\hat h}_{\mu\nu} +
J\,\hat \varphi + J_a\, \hat\eta_a + J^{\mu}{\hat A}_{\mu} \right)
\label{bi}
\end{equation} 
where the currents are
\begin{eqnarray}
&&J_{\mu\nu}(x)= \kappa_{orb} \,M\,\eta_{\mu 0} \,\eta_{\nu 0}\,
\delta^{5}(\vec{x}) ~~~,~~~ J(x)=J_a(x)=
\frac{\kappa_{orb}\,M}{2}\,\delta^5\left(\vec{x}\right) ~~~, \nn\\
&&~~~~~~~~~~~~~~~~~~~~~~~~J_{\mu}(x)= - \sqrt{2}\,\kappa_{orb} \,M\,
\eta_{\mu 0}\, \delta^{5}(\vec{x})~~.
\label{cur}
\end{eqnarray}
The quadratic (and higher order) terms of the boundary action will not
be needed in our calculations since they give rise only to tadpole
diagrams which vanish in dimensional regularization. 

In this theory, the one-point function of a generic bulk field $\hat
\Psi (x)$ is given by
\begin{equation}
\int \left[D\hat h\, D\hat\varphi \,D \hat A\, D\hat\eta\right]
\hat\Psi (x) \, {\rm e}^{{\rm i} S_0 }\,{\rm e}^{{\rm i} S_I }\, {\rm
e}^{{\rm i} S_{\rm boundary}} \equiv  \left< \hat\Psi(x) \,{\rm
e}^{{\rm i} S_I }\, {\rm e}^{{\rm i} S_{\rm boundary}} \right>~~.
\label{dam}
\end{equation}
Expanding the two exponentials, we generate a perturbative series
whose various terms correspond to diagrams that contain a different
number of bulk and boundary interactions. The first two diagrams in
this series are represented in Figure \ref{dil} and describe,
respectively, the leading and next-to-leading terms in the large
distance expansion of the classical bulk fields.  In particular, the
leading contribution is obtained from (\ref{dam}) by neglecting  $S_I$
and expanding at first order the exponential containing $S_{\rm
boundary}$.  Applying this procedure to the dilaton, we find that the
leading contribution to its one-point function is 
\begin{equation}
\hat\varphi^{(1)}(x)={\rm i}\, \int d^{6} y \,\,\Big< \hat\varphi(x)
\,J(y)\,  \hat\varphi(y) \Big> =\frac{\kappa_{orb}\,M}{2}\, \int
\frac{d^{5} k}{(2\pi)^{5}} \frac{{\rm e}^{{\rm i}k\cdot x}}{k^2}~~.
\label{aonep}
\end{equation}
By using the following expression for the Fourier transform  
\begin{equation}
\int \frac{d^{d} k }{(2\pi)^{d}}  \frac{{\rm e}^{{\rm i}k\cdot
x}}{k^{2\alpha}} =  \frac{2\alpha}{2^{2\alpha}\,\pi^{{d}/2}} \frac{
\G\left(\frac{{d}}{2}+1 -\alpha\right)}{\G\left( 1+\alpha\right)}
\,\frac{1}{(d-2\alpha)} \, \frac{1}{|x|^{d-2\alpha}}~~,
\label{gfour}
\end{equation}
we easily see that (\ref{aonep}) becomes
\begin{equation}
\hat\varphi^{(1)}(x)=\frac{1}{ \ky}\,\frac{1}{4} \,\frac{Q}{r^3}
\label{Afirst}
\end{equation}
where  $Q$ is the parameter defined in Eq. (\ref{Q0}). This is
precisely the leading term at large distance of the dilaton produced
by the non-BPS D-particle (see Eq. (\ref{dil20})). In a similar manner
we can compute the asymptotic behavior of the other bulk fields and
find complete agreement with the results reported in
Eq.s (\ref{scala20})-(\ref{met20}).  We now compute the
next-to-leading order of the one-point function (\ref{dam}). This is
obtained by expanding the exponential of $S_I$ at first order and the
exponential of $S_{\rm boundary}$ at second order.
\begin{figure}[ht]
\begin{center}
{\scalebox{1}{\hspace{-2cm}\includegraphics{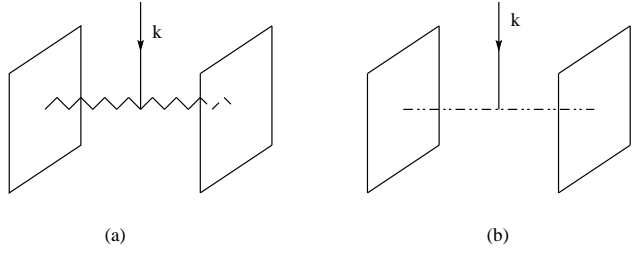}}}
\caption{\small The next to leading order contributions for the
dilaton.  Diagram (a) represents  the two boundary contribution via
graviton exchange while diagram (b)  corresponds to the gauge  field
contribution.}
\label{dilaton}
\end{center}
\end{figure}
Applying this procedure to the dilaton, we find two contributions
corresponding to the diagrams in Figure \ref{dilaton}, so that we can
write
\begin{equation}
\hat\varphi^{(2)}(x) = A^{\varphi AA}_{\varphi}(x) +A^{\varphi
hh}_{\varphi}(x)~~.
\label{avarphi}
\end{equation}
The first term, due to the coupling of a dilaton with two gravitons
(see Figure \ref{dilaton}a),  is equal to \bea A_{\varphi}^{\varphi h
h}(x)&=&{\rm i}^3 \int d^6 y\int d^6 z \int d^6 u 
\label{dam1}  \\
&&\!\!\!\!\!\!\!  \left< \hat\varphi(x)\, \hat\varphi(y)\, J(y)
\,\de_{\mu}\hat\varphi(z)\,  \de_{\nu}\hat\varphi(z) \left({\hat
h}^{\mu\nu} (z) - \frac{1}{2}\,  {\hat h}^{\;\tau}_{\tau}(z) \,
\eta^{\mu\nu}\right) {\hat h}_{\rho\sigma}(u)\, J^{\rho\sigma} (u)
\right>~~. \nn  \eea By performing all contractions and using the
explicit expressions for the propagators, it is not difficult to see
that $A_{\varphi}^{\varphi h h}(x)=0$. Moreover, it is interesting to
notice that this result holds in any space-time dimension.  The second
term in (\ref{avarphi}) corresponds to the diagram of Figure
\ref{dilaton}b that is given by   \bea A_{\varphi}^{\varphi AA}(x)&=&
\frac{{\rm i}\, \kappa_{orb}}{8} \int\! d^6 y\int \!d^6 z \int\! d^6 u
\left< \hat\varphi(x) \,\hat\varphi(y)\,   \hat F^{\,2}(y) \,\hat
A_{\mu}(z) \,J^{\mu}(z) \,\hat A_{\nu}(u)\, J^{\nu}(u) \right> \nn\\
&=& \,\ky^3\,M^2\int\! \frac{ d^{5} k }{(2\pi)^{5}}\, \frac{{\rm
e}^{{\rm i}k\cdot x}}{k^2} \, \int \!\frac{d^{5} p}{(2\pi)^{5}}\, \,
\frac{p \cdot (p+k)}{p^2\, (k+p)^2}~~.
\label{dAA}
\eea  The second integral in (\ref{dAA}) can be easily evaluated with
standard techniques. For the sake of generality we give the result of
the previous  integral for an arbitrary value $d$ of the number of
transverse directions. One gets
\begin{equation}
\int \frac{d^{d} p}{(2\pi)^{d}} \,\frac{p \cdot  (p+k) }{k^2\, p^2
\,(k+p)^2}= -  \frac{1}{2}\,\frac{1}{(4\pi)^{d/2}}\, B\left(
\frac{d}{2}-1\,,\,\frac{d}{2}-1 \right) \,  \Gamma\left(2-\frac{d}{2}
\right) \,(k^2)^{d/2-2}~~.
\label{amsp}
\end{equation}
Inserting this result in (\ref{dAA}) and using (\ref{gfour}) for
$\alpha=-1/2$ and $d=5$, we finally get 
\begin{equation}
\hat\varphi^{(2)}(x)= -\,\frac{1}{\ky}\, \frac{1}{8}\,
\left(\frac{Q}{r^3}\right)^2
\label{amdil}
\end{equation}
which agrees with the next-to-leading term of Eq. (\ref{dil20}).

The same calculations that lead to $A_{\varphi}^{\varphi h h}(x)=0$,
also imply that $A_{\eta}^{\eta h h}(x)=0$. In fact, the gravitational
couplings of the scalar fields $\hat\eta_a$ are the same  as those of
the dilaton, and therefore also this diagram does not contribute.  On
the other hand, since  $\hat\eta_a$ does not couple to any other bulk
field, from the vanishing of $A_{\eta}^{\eta h h}(x)$ we can deduce
that $\hat\eta_a$ does not receive any correction at the next-to-leading 
order. Actually, also the higher orders for these fields are vanishing 
and thus the leading term for $r\rightarrow \infty$ already gives
the exact result (see Eq. (\ref{scala3})).

With this same method we can compute the next-to-leading term for
the gauge field $\hat A_0$. This is the sum of two terms which arise
from the bulk interaction of two gauge fields with a dilaton and a 
graviton respectively, namely
\begin{equation}
{\hat A_0}^{\,(2)} (x) = A_{A}^{A\varphi A} (x)  + A_{A}^{AhA} (x)~~.
\label{rrampl}
\end{equation}
Finally, the next-to-leading term of the graviton  
is produced by the bulk interactions involving three gravitons, 
two dilatons and one graviton, 
two gauge fields and one graviton, and two scalars and one graviton, that is
\begin{equation}
{\hat h}_{\mu \nu}^{\,(2)} (x) = 
A_{\mu\nu}^{hhh}(x)  + A_{\mu\nu}^{h\varphi\varphi} (x) + 
A_{\mu\nu}^{h AA} (x) +A_{\mu\nu}^{h\eta\eta}(x)~~.
\label{hmunu}
\end{equation}
All terms in (\ref{rrampl}) and (\ref{hmunu}) can be computed 
following the procedure outlined before, and after some lengthy algebra
one gets precisely the next-to-leading behavior of the gauge field
and the metric written in Eq.s (\ref{vec20})-(\ref{met20}).

\vskip 1.5cm
\noindent
\appendix{\large{\bf {Appendix B}}}

\renewcommand{\theequation}{B.\arabic{equation}}
\setcounter{equation}{0}
\vskip 0.5cm
\noindent
In this appendix we explicitly derive the non-BPS D-particle solution 
(\ref{dil3})-(\ref{met3}), and the most general one presented in
Eq.s (\ref{solgen})-(\ref{solgen4}).
For the sake of generality we start from a $D$-dimensional action containing 
the metric, the dilaton, the scalars $\eta_a$ and a 
$(p+1)$-form R-R potential with $p<D-3$. The case of the
non-BPS D-particle, considered in Section 2,  
can be obtained by taking in all our equations $p=0$ and 
$D=6$. However, our equations can also be used to derive the 
non-BPS solution in $D=10$ discussed in Section 2.
Actually they can also be 
used for the usual BPS D-branes in ten dimensions. We start from the 
following action
\begin{equation}
S = S_{\rm bulk} + S_{\rm boundary}
\label{bb3}
\end{equation}
where
\begin{equation}
S_{\rm bulk}=\frac{1}{2 \kappa_{orb}^2}\int d^D x \sqrt{-\det G}
\left[ {\cal R}(G) - 
\partial_{\mu}\varphi\, \partial^{\mu}\varphi -\partial_{\mu}\eta_a\,
\partial^{\mu}\eta_a - \frac{1}{2(p+2)!} {\rm e}^{ a\, \varphi}F^2_{p+2}\right]
\label{bulk}
\end{equation}
and
\begin{equation}
S_{\rm boundary} = -M\int d^{p+1} \xi \,
{\rm e}^{-\frac{a}{2}\varphi-\frac{1}{2} 
\sum_{a} \eta_a} 
\sqrt{-\det G_{\alpha\beta}} + M \int A_{p+1}
\label{bound}
\end{equation}
where $\kappa_{orb}$ has been defined in (\ref{con}), while $M= N
M_p$ with
\begin{equation}
M_p=\frac{\sqrt{2}\,T_p}{(2\pi)^2\kappa_{orb}V^{1/2} }~~. 
\label{Mp}
\end{equation}
As mentioned above, we treat simultaneously the
cases of non-BPS branes in both $D=6$ and $D=10$. This can be done
by taking the constant $a$ 
to be given by
\begin{equation}
a = \frac{D-4-2p}{\sqrt{D-2}}~~.
\label{a10}
\end{equation}
Clearly, if $D=10$ in (\ref{Mp}) we have to put
$\kappa_{10}$ in place of $\kappa_{orb}$ and delete the factor of 
$(2\pi)^2 V^{1/2}$. 
Moreover, in the ten dimensional case
there are no scalars $\eta_a$ and in the case of the non-BPS branes there is 
no R-R field. 
Finally for the BPS branes there is no factor of $\sqrt{2}$ in the
brane tension (\ref{Mp}).

By varying the action ({\ref{bb3}), we get the equations of motion for 
the various fields. In particular, we have 
\begin{equation}
\frac{1}{\sqrt{-\det G}}\, \partial_{\mu}\left( \sqrt{-\det G}\, G^{\mu\nu} 
\partial_{\nu} \varphi\right)
-\frac{a}{4}\, \ee^{a \varphi}\,\frac{1}{(p+2)!}\,F^2_{p+2}=
\frac{a}{2}\,T(x)\, \delta^{d}(\vec{x})
\label{scaleq}
\end{equation}
for the dilaton,
\begin{equation}
\frac{1}{\sqrt{-\det G}}\,\partial_{\mu}\left( \sqrt{-\det G}\, G^{\mu\nu}\,
 \partial_{\nu} 
\eta_a\right) =\frac{1}{2} \,T(x)\,\delta^{d}(x)
\label{eqscal}
\end{equation}
for the scalars and
\begin{equation}
\partial_{\mu_1}\left( \sqrt{-\det G}\, \ee^{a\varphi}\,
G^{\mu_1\nu_1}\cdots G^{\mu_{p+2}\nu_{p+2}}\,\frac{ F_{\nu_1\cdots\nu_{p+2}} }
{(p+1)!}\right)=
- 2 \,M \,\kappa_{orb}^2 \,G_0^{\,\,\mu_2}\cdots G_p^{\,\,\mu_{p+2}}\,
\delta^{d}(\vec{x})
\label{rameq}
\end{equation}
for the R-R field.
Finally, the Einstein equations for the metric can be written in a simple
form by first evaluating their trace, and then plugging it back into
the original equations, obtaining 
\begin{eqnarray}
&&\!\!\!\!\!\!\!\!\!\!\!\!\!\!\!\!{\cal R_{\mu\nu}} -
\frac{\ee^{a\varphi}}{2(p+2)!}\left[(p+2) F_{\mu\mu_2\cdots \mu_{p+2} }\,
F_{\nu}^{\,\,\mu_2\cdots\mu_{p+2} } - G_{\mu\nu}\, \frac{(1+p)}{4}\,F^2_{p+2}\right]
\\
&&\!\!\!\!\!\!\!\!\!-\,\partial_{\mu}\varphi\,\partial_{\nu}\varphi 
-\partial_{\mu}\eta_a \,\partial_{\nu} \eta_a = T(x)\,
\left(G_{\mu\alpha}\,G_{\nu\beta} \,G^{\alpha \beta}
-\frac{p+1}{4}\,G_{\mu\nu}\right)\, \delta^{d}
(\vec{x}) \nn
\label{meteq}
\end{eqnarray}
where
\begin{equation}
T(x)= -\,M\,\kappa_{orb}^2\,\ee^{-\frac{a}{2}\varphi -\frac{1}{2}\sum_a\eta_a}
\,\frac{ \sqrt{-\det G_{\alpha\beta}} }{ \sqrt{-\det G} }\quad .
\label{tsor}
\end{equation}

We now solve the previous equations using the following {\it Ansatz}
for the metric
\begin{equation}
ds^2= B^2(r) \,\eta_{\alpha\beta}\, dx^{\alpha}\,dx^{\beta} + F^2(r) 
\,\delta_{ij}\, dx^{i}\,dx^{j} 
\label{metric}
\end{equation}
where $\alpha,\beta =0,...,p$ and $i,j=p+1,...,d\equiv D-p-1$, 
and assuming that all other fields are functions only of 
the radial coordinate $r$. 
Under these assumptions, the dilaton equation (\ref{scaleq}) becomes
\begin{eqnarray}
&&\frac{1}{r^{d-1}}\, \left( r^{d-1}\, B^{p+1}\, F^{d-2}\,  
\varphi' \right)'\,
+\,\frac{a}{4}\,\ee^{a\varphi}\, F^{d-2}\, B^{-p-1}\,\left( 
A_{01\cdots p}'\right)^2\nn\\
&&~~~~~~~~~~~~~~~~~~=\frac{a}{2} \, 
B^{p+1}\,F^d \,T(x)\,\delta^{d}(\vec{x})~~,
\label{rscaleq}
\end{eqnarray}
the scalar equation (\ref{eqscal}) becomes
\begin{equation}
\frac{1}{ r^{d-1} }\,\left( r^{d-1}\, B^{p+1}\, F^{d-2}\,  
\eta_a' \right)'=
\frac{1}{2}\, B^{p+1}\,F^{d} \,T(x) \,\delta^{d}(\vec{x})~~, 
\label{reqscal}
\end{equation}
while R-R field equation (\ref{rameq}) becomes
\begin{equation}
\frac{1}{r^{d-1} }\,\left( 
r^{d-1}\, B^{-p-1}\, F^{d-2}\, \ee^{a\varphi} 
\, A_{01\cdots p}'\right)'= 2\, M\, \kappa^2_{orb} \,\delta^{d}
(\vec{x}) 
\label{rrameq}
\end{equation}
where $'\equiv d/dr$. Finally, from the Einstein equations (\ref{meteq}) 
we get
\begin{eqnarray}
&&F^{-2} \left\{ - \xi'' -\left( \log F\right)'' -\frac{d-1}{r} 
\left(\log F \right)' -(p+1)\left[\left( \log B \right)' \right]^2+ \xi'
 \left(\log F \right)' \right.\nonumber\\
&&- (d-2) \left[\left( \log F\right)' \right]^2\Bigg\}
-F^{-2}\,\left(\varphi'\right)^2 -F^{-2} \,\sum_a \left( \eta_a' 
\right)^2\nonumber\\
&& - \frac{\ee^{a\varphi}}{2}\,\frac{d-2}{D-2}\,F^{-2}\,B^{-2(p+1)}
\left( A_{01\cdots p}'\right)^2 =
-\frac{p+1}{D-2}\, T(x)\, \delta^{d}(\vec{x})
\label{rmetric}
\end{eqnarray}
for the components $R_{r}^{r}$,
\begin{eqnarray}
&&F^{-2} \left\{-(\log B )'' -(\log B)'\left[ \xi'+\frac{d-1}{r}
\right]\right\} \nonumber\\
&&-\frac{\ee^{a\varphi} }{2}\,\left( - \frac{d-2}{D-2}\right)\,F^{-2}\,
B^{-2(p+1)}\,
\left(A_{01\cdots p}'\right)^2 =\frac{d-2}{D-2}\,T(x)\,
\delta^{d}(\vec{x})\nonumber\\
&&\label{raa}
\end{eqnarray}
for the components $R^{\alpha}_{\alpha}$, and   
\begin{eqnarray}
&&F^{-2}\,\left[-(\log F)'' -\frac{d-1}{r}\,(\log F)'-(\log F)'\,\xi'-
\frac{\xi'}{r}\right]
\nonumber\\
&& -\frac{\ee^{a \varphi}}{2}\, \frac{p+1}{D-2}\,B^{-2p-2} \left( 
A_{01\cdots p} '\right)^2 =
-\frac{p+1}{D-2}\,T(x)\,\delta^{d}(\vec{x})
\label{ttr}
\end{eqnarray} 
for the components $R^{\bar{\alpha}}_{\bar{\alpha}}$ where the index 
${\bar{\alpha}}$ corresponds to the angular variables. In these equations we
have introduced the function 
\begin{equation}
\xi =(p+1)\log B+ (d-2) \log F~~.
\label{phi}
\end{equation}
Following for example \cite{Arg}, 
we now multiply Eq. (\ref{raa}) by a factor of
$(p+1)$ and  Eq. (\ref{ttr}) by a factor of $(d-2)$, and then sum the two
expressions. In this way we see that the 
function $\xi$ obeys a simple differential
equation, namely
\begin{equation}
\left[ r^{2d-3} \,\left( \ee^{\xi}\right)'\right]'=0 \quad .
\label{eqphi}
\end{equation}
This is the Laplace equation in $2d-1$ dimensions and its most general solution
can be written as 
\begin{equation}
e^{\xi}= \hat{C} + C \left( \frac{Q_{p}}{r^{d-2}} \right)^2 
\label{solphi}
\end{equation}
where $\hat C$ and $C$ are arbitrary constants, and, for later convenience,
we have introduced the dimensionful quantity
\begin{equation}
Q_p= \frac{2\,\kappa_{orb}^2\, M_p}{(d-2)\,\Omega_{d-1}}~~.
\label{Qp9}
\end{equation}
In order to have an asymptotically 
flat metric, we must choose $\hat C=1$, and thus
we can write
\begin{equation}
\ee^{\xi} \equiv B^{p+1}\,F^{d-2} = 
f_{-} (r) f_{+} (r)
\label{relmet}
\end{equation}
with
\begin{equation}
f_{\pm} (r) = 1 \pm  x \frac{Q_p}{r^{d-2}} \hspace{1cm}; \hspace{1cm}
x^2=-C~~.
\label{f+f-}
\end{equation}
Inserting Eq. (\ref{relmet}) into Eq.s (\ref{rscaleq})-(\ref{ttr}), we get 
\begin{equation}
\frac{\ee^{- \xi}}{r^{d-1}}\, \left( r^{d-1}\, \ee^{\xi} \, 
\varphi'\,\right)'\,+\,
\frac{a}{4}\, 
\ee^{a \varphi} \,B^{-2(p+1)} \,\left(A_{01\cdots p}'\right)^2 =
\frac{a}{2}\,F^2\,T(x) \,\delta^{d}(\vec{x})
\label{nscaleq}
\end{equation}
for the dilaton, 
\begin{eqnarray}
\frac{\ee^{-\xi}}{r^{d-1}} \, \left( r^{d-1}\, \ee^{\xi}\, \eta_a'
\right)' =
\frac{1}{2}\,F^2\,T(x)\,\delta^{d} (\vec{x})
\label{neqscal}
\end{eqnarray}
for the scalars $\eta_a$,
\begin{eqnarray}
&&\frac{1}{r^{d-1}}\, \left( r^{d-1}\, \ee^{a \varphi}\, B^{-(p+1)}\, F^{d-2}
\, A_{01\cdots p}' \right)'= 2\, M\, \kappa_{orb}^2\, \delta^{d}(\vec{x})
\label{nrameq}
\end{eqnarray}
for the R-R field, while Eq. (\ref{raa}) can be rewritten as
\begin{eqnarray}
&&\frac{e^{- \xi}}{r^{d-1}}\,\left[ r^{d-1} \,\ee^{\xi}\, (\log B)'\right]' -
\frac{\ee^{a \varphi}}{2} \,\frac{d-2}{D-2} \,B^{-2(1+p)}\left( 
A_{01\cdots p}'\right)^2 \nn\\
&&~~~~~~~~~~~~~~~~~~~~=
-\frac{d-2}{D-2}\,F^2\,T(x)\,\delta^{d}(\vec{x}) ~~.
\label{nttr}
\end{eqnarray}
Multiplying Eq. (\ref{nscaleq}) by $2 (d-2)/(D-2)$ and Eq. (\ref{nttr}) by 
$a$, and then summing the resulting expressions we get:
\begin{equation}
\left( r^{d-1} \,\ee^{\xi}\,  Y' \right)'=0
\label{2cond}
\end{equation}
where
\begin{equation}
Y \equiv  \frac{D-2}{d-2} \log B + \frac{2}{a}  \varphi~~.
\label{yyy}
\end{equation}
The solution of this equation is
\begin{equation}
\ee^{Y} \equiv B^{(D-2)/(d-2)}\, \ee^{2 \varphi/a} =  \left(
\frac{f_{-} (r)}{f_{+} (r) } \right)^{\epsilon}
\label{soly}
\end{equation}
where $\epsilon$ is an arbitrary integration constant
\footnote{Here and in the following, when 
$a=0$ ({\it e.g.} $p=1$ in $D=6$ and $p=3$ in $D=10$)
the equations are 
ill-defined. However, their general solutions are valid also in these cases.}.
Actually the most 
general solution of Eq. (\ref{2cond}) admits an additional arbitrary constant
which, however, we have fixed by requiring that $Y$ vanish for $r \rightarrow 
\infty$.
Using Eq. (\ref{ttr}) in Eq. (\ref{rmetric}), and 
expressing $B$ and $F$ in terms
of $\xi, Y$ and $\varphi$ we can rewrite Eq. (\ref{rmetric}) as follows
\[
\frac{\ee^{a\varphi}}{2}\, B^{-2(p+1)}
\left( A_{01\cdots p}'\right)^2= \frac{4}{D-2}\,\left[
\frac{a^2 (D-2)}{4} + (p+1)(d-2) \right]
\left( \frac{\varphi'}{a} \right)^2 +
\]
\[
+ \sum_a \left(\eta_a'\right)^2
+ \xi'' - \frac{1}{d-2} \left(\xi' \right)^2 - \frac{\xi'}{r} +
\frac{(d-2)(p+1)}{D-2} \left[ \left( Y'\right)^2 - Y' 
\left(\frac{4 \varphi'}{a}\right) \right]~~.
\]
\begin{equation}
\label{Feq}
\end{equation}
Let us start by examining the case of a non-BPS D-brane in $D=6$. 
When we use Eq. (\ref{Feq}) in 
the dilaton equation (\ref{nscaleq}), the latter becomes
\[
\frac{\ee^{- \xi - 2 \varphi/(1-p) + (d-2)(p+1) Y/2}}{r^{d-1}}\,
\left[ r^{d-1} \,\ee^{\xi - (d-2)(p+1) Y/2 } \,\left( {\ee}^{\frac{2}{1-p} 
\varphi}\right)'\right]' + \sum_a\left(  \eta_a'\right)^2   
\]
\begin{equation}
 - \frac{d-1}{d-2} \,( \xi ' )^2 -
2(d-1)\, \frac{\xi'}{r} +  \frac{(d-2)(p+1)}{4} 
\left( Y'  \right)^2 =F^2\,T(x)\,\delta^{(5-p)}(\vec{x})~~.
\label{dileq1}
\end{equation} 
Using  Eq.s (\ref{relmet}) and (\ref{soly}) with $D=6$  
for the functions $\xi$ and $Y$,  and the fact that the scalar 
fields $\eta_a$ 
satisfy the same homogeneous equation as $Y$ and therefore are given by
\begin{equation}
\ee^{\eta_a} = \left( \frac{f_{-} (r)}{ f_{+} (r) }\right)^{\delta}~~,
\label{eta23}
\end{equation} 
one can see that the dilaton equation (\ref{dileq1}) becomes
\begin{eqnarray}
&&\!\!\!\!\!\!\!\!\!\!\!
\frac{\ee^{- \xi - 2 \varphi/(1-p) + (d-2)(p+1) Y/2}}{r^{d-1}} \,
\left[ r^{d-1}\, \ee^{\xi - (d-2)(p+1) Y/2 } \left( \ee^{\frac{2}{1-p} 
\varphi}\right)'\right]' +
C\,\frac{d-2}{r^2}\,\left( \frac{Q_p}{r^{d-2}} \right)^2\, \ee^{-2 \xi} 
\nonumber\\
&&\times
\left[ - 16 (d-2) \delta^2 + 4 (d-1) - (d-2)^2 (p+1) \epsilon^2  \right] =
F^2\,T(x)\,\delta^{(5-p)}(\vec{x})~~.\nonumber\\
&&\label{dileq}
\end{eqnarray}
This must be considered together with the equation (\ref{nrameq})
for the R-R field, which becomes
\begin{equation}
\frac{1}{r^{d-1}} \,\left( r^{d-1}\, \ee^{\xi -(p+1)(3-p) Y/2
+\frac{4}{1-p}\varphi} \, A_{01\cdots p}'
\right)=2\, M \,\kappa_{orb}^2\, \delta^{(5-p)}(\vec{x})~~.
\label{rreq}
\end{equation}
Finally, using Eq.s (\ref{yyy}) and (\ref{2cond}),  Eq. (\ref{nttr}) 
can be rewritten as follows
\begin{equation}
\frac{\ee^{- \xi}}{r^{d-1}} \, \left[ r^{d-1}\, \ee^{\xi}\, 
\left(\frac{2 \varphi'}{1-p} \right) \right]' +\frac{1}{2} 
\,\ee^{4 \varphi/(1-p) - (p+1)(d-2) Y/2} 
\,\left( A_{01\cdots p}'\right)^2= - T(x)\,
\delta^{(5-p)}(\vec{x})~~.
\label{raa3}
\end{equation}
In the following we want to find the most general solution of 
Eq.s (\ref{Feq}) and (\ref{dileq})-(\ref{raa3}) excluding the origin where
the boundary action is located 
and corresponding to vanishing
values of $\varphi$ and $A_{01 \dots p}$ for $r \rightarrow \infty$.
Under these conditions, we find that 
(\ref{dileq}) is solved  by
\begin{equation}
\ee^{2 \varphi/(1-p)} = \left( \frac{\cosh X(r) + \gamma \sinh X(r)}{\cosh \alpha 
+\gamma \sinh \alpha} \right) 
\left(\frac{f_{-} (r)}{f_{+} (r)} \right)^{(d-2)(1+p) \epsilon/4}
\label{dil78}
\end{equation}
where
\begin{equation}
X(r) = \alpha + \beta \log \frac{f_{-} (r)}{f_{+} (r)}~~,
\label{X34}
\end{equation}
provided that the following relation is satisfied
\begin{eqnarray}
- \,4 (d-2) (x\beta)^2 -16 (d-2) (x\delta)^2 + 4 (d-1)
\,x^2 ~~~~~~~~~~& & ~~~~~~~\nonumber\\
+\; (d-2)^2 (p+1) (x\epsilon)^2
\left(\frac{(d-2)(p+1) }{4} -1 \right) =  0 ~~.& & 
\label{eq241}
\end{eqnarray}
Inserting the solution for the dilaton into Eq. (\ref{rreq}) 
and neglecting again the source term, we find that the R-R field is given by
\begin{equation}
A_{01 \dots p} =  \sqrt{2(\gamma^{2} - 1)}   
\left[ \frac{ \sinh X(r)(\cosh \alpha + \gamma \sinh \alpha) }{ 
\cosh X(r) + \gamma \sinh X(r)} - \sinh \alpha\right]
\label{rr45}
\end{equation}
where the overall constant has been determined in terms of $\gamma$ through  
Eq. (\ref{raa3}).

Finally from Eq.s (\ref{relmet}), (\ref{soly}) and  (\ref{dil78}), one
can find the explicit expressions for the components of the metric:
\begin{equation}
B^2 = \left( \frac{\cosh X(r) + \gamma \sinh X(r)}{\cosh \alpha 
+ \gamma \sinh \alpha}
 \right)^{- (d-2)/2} 
\left(\frac{f_{-} (r)}{f_{+} (r)} \right)^{\epsilon (d-2)/2[ 1 - (d-2)(p+1)/4]}
\label{bbb}
\end{equation}
and
\begin{eqnarray}
F^2 &=& \left( \frac{\cosh X(r) + \gamma \sinh X(r)}{\cosh \alpha 
+ \gamma \sinh \alpha}
 \right)^{ (p+1)/2} \left( f_{-} (r) f_{+} (r) \right)^{\frac{2}{(d-2)}} 
\nn\\
&&\times\left( \frac{f_{-} (r)}{ f_{+} (r)} \right)^{- \epsilon 
(p+1)/2 [1 - (d-2)(p+1)/4 ]}~~. 
\label{fff}
\end{eqnarray}
Eq.s (\ref{eta23}), (\ref{dil78}) and (\ref{rr45})-(\ref{fff}) represent
the most general solution of the field equations derived from the 
action (\ref{bulk}) which describe a static, 
spherically symmetric configuration, with asymptotically flat geometry 
and vanishing v.e.v.'s at infinity for the gauge 
and scalar fields. The solution depends on five arbitrary 
parameters $\alpha$, $\beta$, $\gamma$, $\delta$ and $x$  
($\epsilon$ is in fact determined 
in terms of the others through Eq. (\ref{eq241})). 
Setting $p=0$ one recovers the
solution presented in Section 5 (see Eq.s (\ref{solgen})-(\ref{solgen4})).

In writing this solution, we have not used the precise form of the
source terms, or equivalently we have not imposed that the behavior 
of the various fields at infinity be consistent with what follows from 
the boundary state, which is given by 
\begin{equation}
\varphi \simeq \frac{1-p}{4} \,\frac{Q_p}{r^{d-2}}\hspace{1cm}, \hspace{1cm}
\eta_a \simeq \frac{1}{4}\,\frac{Q_p}{r^{d-2}}\hspace{1cm},\hspace{1cm}
A_{01 \dots p} \simeq - \frac{Q_p}{r^{d-2}}
\label{phieta}
\end{equation}
and
\begin{equation}
G_{\alpha \beta} \simeq \eta_{\alpha \beta}
\left(1 - \frac{d-2}{4} \,\frac{Q_p}{r^{d-2}}\right)  
\hspace{1cm},\hspace{1cm}
G_{ij} \simeq \delta_{ij}\,\left(1 
+ \frac{p+1}{4}\, \frac{Q_p}{r^{d-2}}\right)~~.
\label{met93}
\end{equation}
If we  impose that  our general solution behaves for large $r$ as required by
the previous conditions, we must choose the integration
constants as follows
\begin{equation}
x\,\beta= \pm \frac{\rm i}{4} 
\quad,\quad \gamma= \frac{-\sinh 2\alpha \pm \ii}{\cosh 2\alpha}
\quad,\quad  x \delta= -\frac{1}{8} \quad  
{\rm and}\quad x\to 0 ~~,
\label{fixco}
\end{equation}
with $\alpha$ arbitrary.
For this choice, it is not difficult to see that the $\alpha$ dependence
drops out and the solution is given by
\begin{eqnarray}
\eta_a&=&\frac{1}{4} \,\frac{Q_p}{r^{3-p}} 
\\
\ee^{2 \varphi}&=&\left( 1+\sin \frac{ Q_p}{r^{3-p}}\right)^{\frac{1-p}{2}}
\\
F^2&=&\left(1+\sin \frac{ Q_p}{r^{3-p}}\right)^{\frac{1+p}{4}}\\
B^2&=&\left(1+\sin \frac{ Q_p}{r^{3-p}}\right)^{\frac{p-3}{4}} \\
A_{01\cdots p}&=& -1 +
\frac{ \cos \frac{Q_p}{r^{3-p}} }{1+\sin \frac{Q_p}{r^{3-p}}}~~.
\label{rrsol}
\end{eqnarray}
For $p=0$ this is precisely the solution (\ref{dil3})-(\ref{met3}). 

In the final part of this appendix we use the general equations we have derived
to find the solution corresponding to the non-BPS branes in $D=10$
discussed in Section 2. In this case we have to switch off the scalar fields
$\eta_a$ and the R-R field. Keeping this in mind, 
the dilaton equation  (\ref{nscaleq}) becomes
\begin{equation}
\frac{\ee^{ - \xi}}{r^{d-1}}\, \left(r^{d-1}\,\ee^{\xi}\, \varphi'
 \right)' = \frac{a}{2}\,F^2\,T(x)\, \delta^{(9-p)} (x) ~~,
\label{dil58}
\end{equation}
whereas the metric equations (\ref{nttr}) and (\ref{ttr}) become respectively
\begin{equation}
\frac{\ee^{ - \xi}}{r^{d-1}} \left[r^{d-1}\,\ee^{\xi}\,( \log B)'
 \right]' = - \frac{7-p}{8} \,F^2\,T(x)\,\delta^{(9-p)} (x)
\label{B97}
\end{equation}
and
\begin{equation}
\frac{\ee^{ - \xi}}{r^{d-1}} \, \left[r^{d-1}\,\ee^{\xi}\, ( \log F)'
 \right] '+ \frac{\xi '}{r} =  \frac{p+1}{8}\,F^2\,T(x) \delta^{(9-p)} (x)~~.
\label{F97}
\end{equation}
Finally, Eq. (\ref{Feq}) becomes
\begin{equation}
- \xi'' + \frac{(\xi')^2}{d-2} + \frac{\xi'}{r} - \frac{(p+1)(d-2)}{8} \left[
(Y' )^2 - 4\, Y' \, \frac{ \varphi '}{a} \right] = 2 \left(
\frac{ 2 \varphi'}{a} \right)^2 \quad .
\label{Feq25}
\end{equation}
If we use  Eq.s (\ref{phi}) and  (\ref{yyy}), we can easily see that they 
coincide with (\ref{equations}), provided that the boundary term is omitted.
 
Neglecting for a moment the origin, where the boundary term is located, the 
most general solution of Eq.s (\ref{dil58}) and (\ref{B97}) is given by
\begin{equation}
\ee^{\varphi} = \left(\frac{f_- (r)}{f_+ (r)} \right)^\nu
\hspace{1cm},\hspace{1cm}
B^2 = \left(\frac{f_- (r)}{f_+ (r)} \right)^{\lambda}
\label{sol964}
\end{equation}
where $\lambda$ and $\nu$ are constants to be determined and $f_-$ and $f_+$
are given in Eq. (\ref{f+f-}) with the substitution of $Q_p$ with $\hat Q_p$ 
(the ten dimensional non-BPS D-brane charge 
defined in Eq. (\ref{qp})). From the previous
equations and Eq. (\ref{soly}) we get
\begin{equation}
\epsilon = \frac{4}{7-p} {\lambda} + \frac{2}{a} \nu \quad .
\label{eq654}
\end{equation}
Inserting in Eq. (\ref{Feq25}) the Eq.s (\ref{relmet}), (\ref{soly}) and the 
first equation in (\ref{sol964}) one gets
\begin{equation}
8 \left(\frac{8-p}{7-p} \right) - (p+1)(7-p)\left( \epsilon - 2 \frac{\nu}{a} 
\right)^2 =
8 \,\nu^2 \quad .
\label{equa12}
\end{equation}
Finally imposing that the solution matches also the boundary term we get
\begin{equation}
x = \frac{p-3}{8 \sqrt{2} \nu} = \frac{7-p}{16 \lambda} \quad .
\label{rela34}
\end{equation}
Eq.s (\ref{equa12}) and (\ref{rela34}) imply that
\begin{equation}
\epsilon=0 \hspace{0.7cm},\hspace{0.7cm} x = \sqrt{\frac{7-p}{8(8-p)}}
\hspace{0,7cm},\hspace{0.7cm} {\lambda} = \frac{7-p}{16x}
\hspace{0.7cm},\hspace{0.7cm} \nu = \frac{p-3}{8 \sqrt{2}x } \quad .
\label{solu65}
\end{equation}
Taking into account that the kinetic term for the dilaton  in 
Eq.s (\ref{sbulk})
and (\ref{bulk}) have a factor $2$ of difference in the normalization we see
that Eq.s (\ref{solu65}) reproduce the solution for the ten dimensional non-BPS
D-branes given in Eq.s (\ref{solution}) and (\ref{parameters}).


\end{document}